\documentclass[aps,longbibliography,showpacs,twocolumn,superscriptaddress,amsmath,amssymb,verbatim]{revtex4-2}

\usepackage[LGR,T1]{fontenc}
\usepackage{mathptmx}
\usepackage[utf8]{inputenc}
\usepackage{color}
\usepackage{babel}
\usepackage{textcomp}
\usepackage{multirow}
\usepackage{graphicx}
\usepackage{amsmath}
\usepackage{soul}
\setstcolor{red}
\usepackage{mathtools}
\usepackage[unicode=true,pdfusetitle,
bookmarks=true,bookmarksnumbered=true,bookmarksopen=false,
breaklinks=true,pdfborder={0 0 0},pdfborderstyle={},backref=false,colorlinks=true]
{hyperref}
\hypersetup{
	linkcolor=blue, citecolor=blue,  urlcolor=blue}
\mathchardef\mhyphen="2D 
\def\tcb/{\textcolor{blue}}
\def\half/{$\frac{1}{2}$}
\def\yb3+/{Yb$^\mathrm{3+}$}
\def\jeff/{$J_{\mathrm{eff}}$}
\def\scto/{Sr$_3$CuTa$_2$O$_9$}
\def\ybsc/{Yb$_3$Sc$_2$Ga$_3$O$_{12}$}
\def\mnsn/{MnSnB$_2$O$_6$}
\def\mn2+/{Mn$^{2+}$}
\def\zncu/{ZnCu$_3$(OH)$_6$Cl$_{12}$}
\def\ybmg/{YbMgGaO$_4$}
\def\rucl/{$\alpha$-RuCl$_3$}
\def\musr/{$\mu$SR}

\DeclareUnicodeCharacter{2212}{-}

\begin{document}
\preprint{APS/123-QED}
\title{Magnetic properties of a spin-orbit entangled \jeff/ $=1/2$ three-dimensional frustrated rare-earth hyperkagome}

\author{B. Sana}
\affiliation{Department of Physics\unskip, Indian Institute of Technology Madras\unskip, Chennai\unskip, 600036, India\\ }
\author{M. Barik}
\affiliation{Department of Physics\unskip, Indian Institute of Technology Madras\unskip, Chennai\unskip, 600036, India\\ }
\author{M. Pregelj}
\affiliation{Jo\v{z}ef Stefan Institute, Jamova cesta 39, 1000 Ljubljana, Slovenia}
\affiliation{Faculty of Mathematics and Physics, University of Ljubljana, Jadranska u. 19, 1000 Ljubljana, Slovenia}

\author{U. Jena}
\affiliation{Department of Physics\unskip, Indian Institute of Technology Madras\unskip, Chennai\unskip, 600036, India\\ }
\author{M. Baenitz}
\affiliation{Max Planck Institute for Chemical Physics of Solids\unskip, Nöthnitzer Strasse 40\unskip, 01187 Dresden\unskip, Germany}
\author{J. Sichelschmidt}
\affiliation{Max Planck Institute for Chemical Physics of Solids\unskip, Nöthnitzer Strasse 40\unskip, 01187 Dresden\unskip, Germany}
\author{K. Sethupathi}
\affiliation{Department of Physics\unskip, Indian Institute of Technology Madras\unskip, Chennai\unskip, 600036, India\\ }
\affiliation{Quantum Centre of Excellence for Diamond and Emergent Materials\unskip, Indian Institute of Technology Madras, Chennai\unskip, 600036, India}
\author{P. Khuntia}
\email{pkhuntia@iitm.ac.in}
\affiliation{Department of Physics\unskip, Indian Institute of Technology Madras\unskip, Chennai\unskip, 600036, India\\ }
\affiliation{Quantum Centre of Excellence for Diamond and Emergent Materials\unskip, Indian Institute of Technology Madras, Chennai\unskip, 600036, India}

\begin{abstract}
The  interplay between competing  degrees of freedom can stabilize non-trivial magnetic states in correlated electron materials. Frustration induced strong quantum fluctuations can evade long-range magnetic ordering leading to exotic quantum states  such as spin liquids in two-dimensional spin-lattices such as triangular and  kagome structures. However, the experimental realization of dynamic and correlated quantum states is rare in three-dimensional (3D)  frustrated magnets  wherein quantum fluctuations are less prominent. Herein, we report the crystal structure,  magnetic susceptibility, electron spin resonance and specific heat studies accompanied by crystal electric field calculations on a 3D frustrated magnet \ybsc/. In this material, Yb$^{3+}$ ions form a three-dimensional network of corner-sharing triangles known as hyperkagome lattice without any detectable anti-site disorder. Our thermodynamic results reveal a low-energy state with \jeff/ $=1/2$ degrees of freedom in the Kramers doublet state. The zero field cooled and field cooled magnetic susceptibility taken in 0.001 T rules out the presence of spin-freezing down to 1.8 K. The Curie-Weiss fit of magnetic susceptibility at low temperature yields a small and  negative Curie-Weiss temperature   indicating the presence of a weak antiferromagnetic interaction between \jeff/ $=1/2$ (\yb3+/) moments. The Yb-electron spin resonance displays a broad line of asymmetric shape consistent with the presence of considerable magnetic anisotropy  in \ybsc/. The crystal electric field calculations  suggest that the ground state is well separated from the excited states, which are in good  agreement with experimental results. The absence of long-range magnetic ordering inferred from specific heat data indicates  a dynamic liquid-like ground state at least down to 130 mK. Furthermore, zero field specific heat shows a broad maximum around 200 mK suggesting the presence  of short-range spin correlations in this three-dimensional frustrated  antiferromagnet.

\end{abstract}

\maketitle
\section{Introduction} 
Geometrically  frustrated magnets have drawn significant attention due to their emergent magnetic phenomena arising from competing interactions and quantum fluctuations \cite{doi:10.1142/5697,10.1007/978-3-642-10589-0,Savary_2016,RevModPhys.89.025003}. Two-dimensional (2D) frustrated magnets are prime contenders to host exotic quantum states with fractional quantum numbers  following  the seminal proposal of quantum spin liquid state on a triangular lattice antiferromagnet by P. W. Anderson \cite{ANDERSON1973153}. Frustrated  magnets with next nearest-neighbor exchange interaction, bond dependent Ising interactions and magnetic anisotropy are ideal candidates  in this context \cite{Balents2010,10.1007/978-3-642-10589-0}. A quantum spin liquid (QSL) is characterized by the absence of  long-range magnetic order despite strong
exchange interaction between spins, the absence of local order parameter and spins not freezing down to absolute zero. The spins maintain a highly entangled state in the QSL state, which harbors exotic fractional excitations such as spinons and Majorana fermions that are essential ingredients for future quantum information processing \cite{Balents2010,Savary_2016,RevModPhys.89.025003,Do2017,Banerjee2016,doi:10.1126/science.aay0668}. Geometrically frustrated  triangular, kagome and  hyperkagome lattices,  wherein frustration induced strong quantum fluctuations are at play, offer a promising platform for the experimental realization of QSL. The  identification of  associated non-trivial quasi-particles and their interactions in the QSL state is an attractive track in quantum condensed matter. The  exotic excitations in the QSL state are completely different from spin-wave excitations  observed in conventional magnets with static order \cite{Balents2010,doi:10.1142/5697,10.1007/978-3-642-10589-0}. There are reports on a quite good number of promising two-dimensional quantum spin liquid candidates, however a very few 3D quantum spin liquids are studied so far \cite{PhysRevLett.91.107001,PhysRevB.77.104413,PhysRevLett.106.147204,PhysRevB.93.140408,doi:10.1021/ja053891p,PhysRevLett.100.087202,Khuntia2020, Li2015,PhysRevLett.118.107202,PhysRevLett.122.137201,PhysRevX.8.031028, Banerjee2016,Do2017,Takagi2019,Clark2019,Arh2022,Arh2022,PhysRevB.90.035141,PhysRevLett.116.107203,Chillal2020,Balz2016,PhysRevB.95.174414,Plumb2019,Gao2019,PhysRevLett.122.187201}.

 Remarkably, an excellent three-dimensional lattice to realize exotic magnetic ground state is rare-earth pyrochlore oxides, R$_2$B$_2$O$_7$, where  R$^{3+}$ and B$^{4+}$  are generally trivalent rare-earth and tetravalent transition-metal ions, respectively. In this  rare-earth magnet family, the ground state of some Kramers ion is a well-separated crystal electric field doublet with interacting  \jeff/ $=1/2$ pseudo-spins at low temperatures. Depending upon how the ground state doublet transforms under time-reversal symmetry and local D$_\mathrm{3d}$ crystal field symmetry of the rare-earth site, this pyrochlore can harbor dipole-octupole ground state doublet, which supports exotic  U(1) quantum spin liquid ground state \cite{PhysRevB.95.041106,PhysRevX.12.021015}. For example, in the quantum spin liquid candidate Ce$_2$Zr$_2$O$_7$, $x$ and $z$ components of pseudo-spin-1/2 degrees of freedom of Ce$^{3+}$ ion behave as dipoles, whereas only the $y$ component behaves as an octupole under the time-reversal symmetry and lattice symmetry \cite{PhysRevX.12.021015}. Coupling between  magnetic field and spinons enables to control spinon excitations by applying an external magnetic field  in this octupolar spin liquid candidate \cite{PhysRevB.95.041106}. Among 3D spin  liquid candidates, three-dimensional network of corner-sharing triangles known as hyperkagome lattice has recently drawn significant attention due to their rich  ground state properties. One exemplar is PbCuTe$_2$O$_6$, wherein  spin $S=1/2$ transition metal Cu$^{2+}$ ions constitute a hyperkagome lattice. The local probe techniques  nuclear magnetic resonance (NMR) and muon spin resonance ($\mu$SR) revealed a dynamic ground state down to 20 mK in this hyperkagome \cite{PhysRevLett.116.107203,PhysRevB.90.035141}.  Diffusive continua in magnetic excitation spectra revealed by neutron scattering experiment provides a strong evidence of fractional spinon excitations in this 3D frustrated magnet \cite{Chillal2020}. Another geometrically frustrated effective spin \jeff/ $=1/2$ 3D spin liquid candidate is Na$_4$Ir$_3$O$_8$ \cite{PhysRevLett.99.137207,PhysRevLett.101.197201}. Despite  strong antiferromagnetic  interaction between Ir$^{4+}$ (\jeff/ $=1/2$) moments, which was confirmed by a large Curie-Weiss temperature of $-650$ K, there is no sign of magnetic ordering down to a few  Kelvin. A theoretical study proposed the existence of a Z$_2$ quantum spin liquid state in the quantum regime  (i.e, small spin) of hyperkagome lattice Na$_4$Ir$_3$O$_8$ \cite{PhysRevLett.100.227201}. However, a small splitting of zero field cooled and field cooled magnetization data suggests  the role of  static moments  below 7 K, which  imposes a strong constraint for the unambiguous identification of  spin liquid ground state in this material  \cite{PhysRevLett.115.047201}. The rare realization of 3D QSL is  due to the fact that the  Ne\'el order or spin freezing is energetically favorable  over the QSL state in the case of a 3D frustrated spin-lattice for $S>1/2$. However, frustration induced  strong quantum fluctuations in the $S=1/2$ or $J_\mathrm{eff} =1/2$ system makes it a potential candidate to host a dynamic ground state \cite{ANDERSON1973153,khomskii_2014,Gao2019}. It is highly relevant to  explore whether the disorder interaction or exchange anisotropy or lattice imperfections account for spin-freezing in this type of 3D spin liquid candidates.

In this context, synthesis and investigation of new three-dimensional frustrated spin-lattices, wherein interplay between  competing degrees of freedom and  spin correlations  could lead to exotic quantum states, are highly needed. Considerable efforts have been devoted  in this direction, for instance, disorder-free  4$f$-based hyperkagome lattice, wherein spin frustration and anisotropic interactions governed by spin-orbit interaction stabilizing an effective low energy \jeff/ $=1/2$ ground state, offers an alternate route for the realization of elusive 3D spin-liquids. Hyperkagome lattice-based material Li$_3$Yb$_3$Te$_2$O$_{12}$ wherein Yb$^{3+}$ spins decorated on a corner-shared frustrated  triangular network shows a dynamical ground state down to 38 mK temperature \cite{PhysRevB.106.104404}. Magnetic Yb$^{3+}$ ions show short-range spin correlations with \jeff/ $=1/2$ degrees of freedom in the ground state.  One promising family of compound to  study rare-earth hyperkagome lattice is lanthanide garnets with general formula R$_3$A$_2$X$_3$O$_{12}$, R = rare-earth elements; A = Ga, Sc, In, Te; X = Ga, Al, Li \cite{Mukherjee_2017}. In this series, rare-earth ions occupying two interpenetrating frustrated hyperkagome networks have a rich potential to  harbor myriads of novel physical phenomena. For example, antiferromagnetic spin correlation and local magnetic anisotropy lead to a non-trivial magnetic structure with long-range hidden order state in Gd$_3$Ga$_5$O$_{12}$ \cite{doi:10.1126/science.aaa5326,PhysRevB.104.054440}. Spin clusters behave like a single object due to strong spin correlation and these spin clusters show long-range non-dipolar order although the individual spin  maintains a dynamic liquid-like state \cite{Machida2010,doi:10.1126/science.aaa5326}.  On the other hand, emergent magnetic behavior with short-range spin correlations was observed in \jeff/ $=1/2$ \yb3+/ containing garnet Yb$_3$Ga$_5$O$_{12}$ \cite{PhysRevB.104.064425}. The specific heat data show a $\lambda$-like anomaly which was attributed to long-range magnetic order \cite{PhysRevLett.91.167201}. However, muon spin resonance and M\"ossbauer spectroscopy  measurements indicate the absence of long-range magnetic order \cite{PhysRevLett.91.167201,Hodges}. In an opposite scenario, antiferromagnetically ordered state develops below in the sub-Kelvin temperature range in other members of the garnet family \cite{PhysRevB.100.184415,PhysRevB.105.014441}. Though there are several studies on gallium garnets, magnetic properties of scandium gallium garnets have yet to be investigated so far. Recent study on R$_3$Sc$_2$Ga$_3$O$_{12}$ (R = Tb, Dy, Ho) reveals that most of these garnets undergo a long-range antiferromagnetic ordering state at low temperature \cite{Mukherjee_2017}. The magnetic properties of this family of materials is highly sensitive to external stimuli such as chemical pressure, temperature, or an applied magnetic field. It is pertinent to test the effect of external perturbations such as non-magnetic or magnetic ions on the anisotropy, exchange interactions and hence the underlying  spin Hamiltonian in this interesting class of three-dimensional frustrated magnets. The magnetic moment of \yb3+/ ion is relatively weak compared to other R$^{3+}$ ions. Being a Kramers ion \yb3+/, a combination of crystal electric field and spin-orbit coupling could lead to a  low energy \jeff/ $=1/2$ state  in  the Yb variant of this promising garnet series. This suggests that  \ybsc/ is of special interest to look for  spin-orbit driven ground state.

Herein, we report the synthesis, magnetization, electron spin resonance (ESR), and thermodynamic studies on a  garnet \ybsc/, wherein \yb3+/ ions constitute a three-dimensional frustrated corner-sharing network of triangles, namely, a hyperkagome spin-lattice.  The combination of crystal electric field and spin-orbit coupling stabilizes a low energy pseudo-spin \jeff/ $=1/2$ state of the Kramers doublets of Yb$^{3+}$ ion at low temperature.  The absence of ZFC-FC splitting of magnetic susceptibility in 0.001 T rules out spin-freezing down to 1.8 K in this material. The low-temperature Curie-Weiss fit yields a weak antiferromagnetic interaction between \yb3+/ (\jeff/ $=1/2$) moments in the spin-lattice. The magnetic ground state is investigated via thermodynamic measurements which reveal that this garnet remains in a quantum disordered ground state at least down to 130 mK.  The specific heat in zero field shows a broad maximum around 200 mK implying the presence of short-range spin correlations in this three-dimensional frustrated spin-lattice. The crystal electric field calculations infer that the lowest Kramers ground state is well separated from the excited states and imply considerable magnetic anisotropy which is consistent with experiments.

\section{Experimental details}
Polycrystalline sample of \ybsc/  was prepared via the sol-gel method. Rare-earth oxide Yb$_2$O$_3$ was preheated at 900$^\circ$C  for 6 hours  prior to use  to remove moisture and carbonates. Stoichiometric amount of Yb$_2$O$_3$ (Alfa Aesar, 99.998\%), Sc$_2$O$_3$ (Alfa Aesar, 99.9\%) and Ga$_2$O$_3$ (Alfa Aesar, 99.999\%) were dissolved in hot nitric acid separately in three beakers. The concentration  of these  acidic solutions was reduced by adding and evaporating deionized water. These three nitrate solutions were then mixed together and polyethylene glycol was added to the solution. This solution was kept in between 90$^\circ$C to 120$^\circ$C on a magnetic stirrer for several hours. The gel was dried yielding ash-like powders, which were preheated at 800$^\circ$C for 6 hours, and finally a white color polycrystalline sample was obtained. This is followed by the heat treatment of the resulting sample at 1000$^\circ$C, 1100$^\circ$C and 1200$^\circ$C for a few days. Before each heat treatment, the sample was grounded and pelletized to ensure better homogeneity. Rigaku x-ray diffractometer  was deployed to check the phase purity at room temperature using Cu $K_{\alpha}$ radiation. Quantum Design, SQUID-VSM  was used to perform  magnetization measurements in the temperature range 1.8 K $\leq T \leq$ 350 K under magnetic field 0 T $\leq \mu_0H \leq$ 5 T. Electron spin resonance (ESR) experiments were performed at 9.4 GHz (X-band frequency) on a high quality polycrystalline sample of \ybsc/ material for temperatures down to 4 K. Specific heat measurements were carried  out with a Quantum Design, Physical Properties Measurement System (QD, PPMS) in  magnetic fields up to 7 T and in the temperature range 1.8 K $\leq T \leq$ 200 K. The low-temperature
specific heat  measurement down to 130  mK was performed
using a dilution refrigerator set up attached to
the Dynacool PPMS from Quantum Design, USA. Fitting and modeling of the crystal-electric-field (CEF) effects were preformed using PHI software\cite{chilton2013phi}.
\begin{figure}[ht]
	\includegraphics[width=8cm, height=12cm]{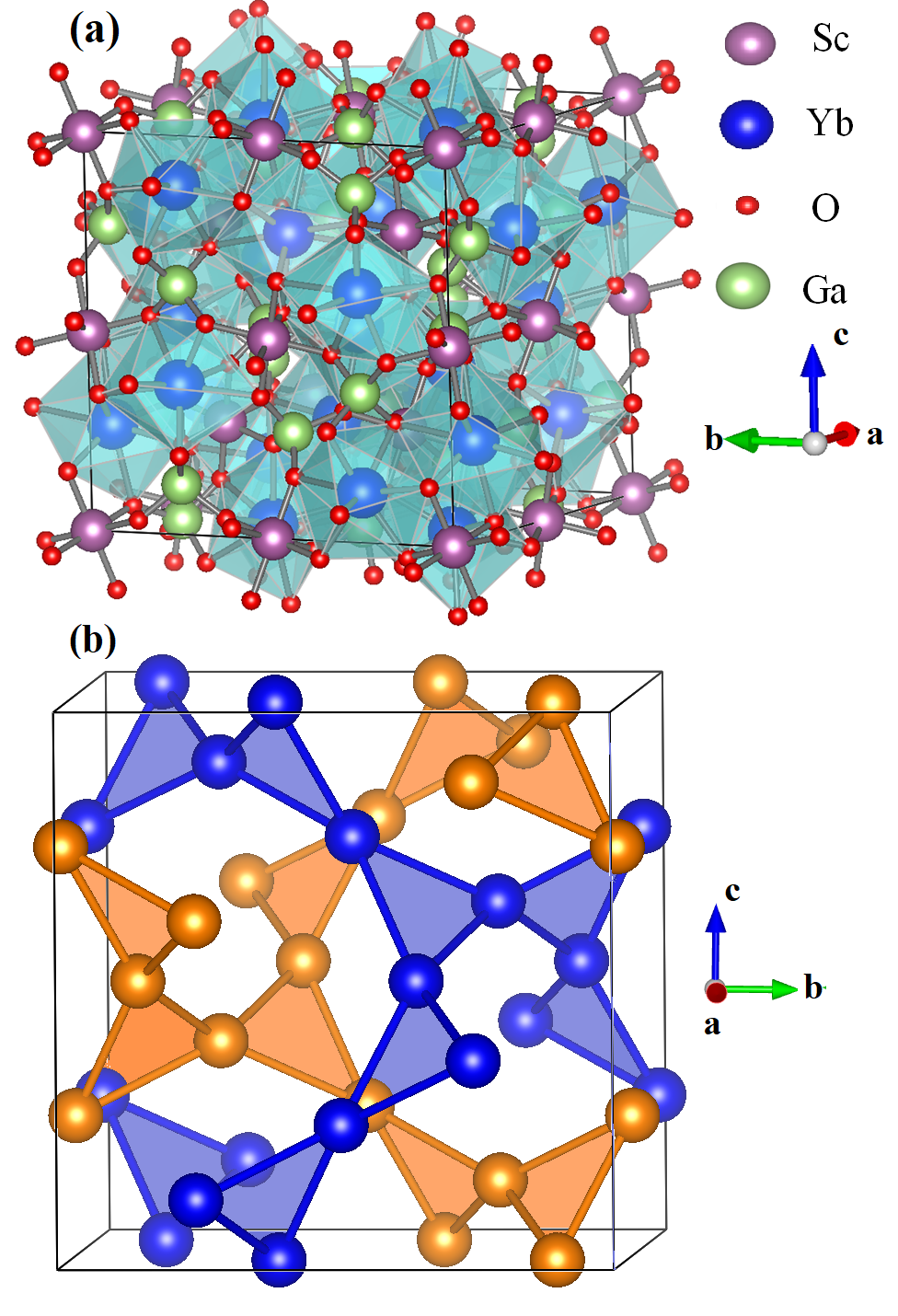}
	\caption{\textbf{(a)} Crystal structure of \ybsc/. \textbf{(b)} Two interpenetrating   hyperkagome network of Yb$^{3+}$ ions.}{\label{structure}}
\end{figure}

\begin{figure}[ht]
	\includegraphics[width=9cm, height=7.5cm]{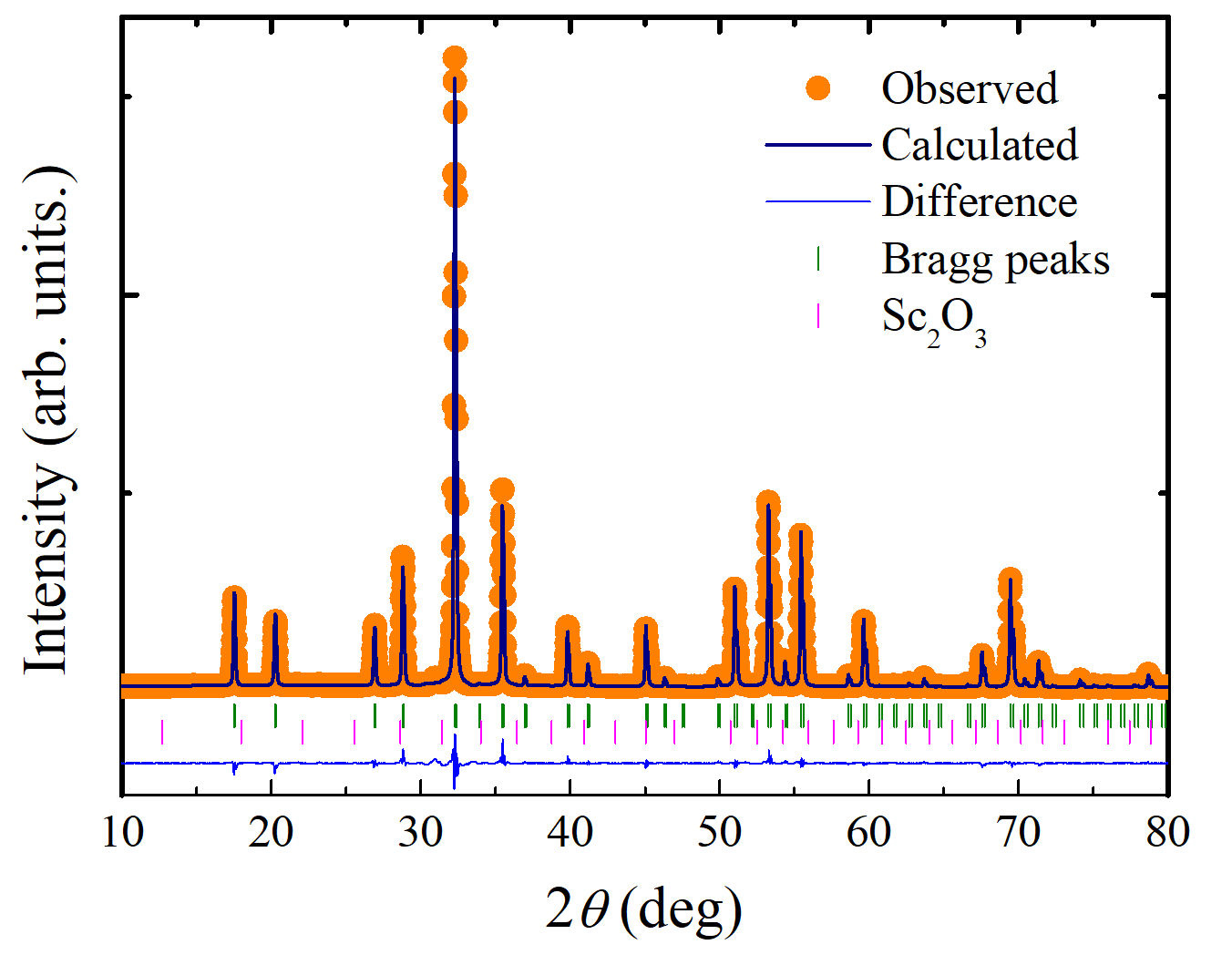}
	\caption{Rietveld refinement of powder XRD data taken at room temperature.}{\label{fig:xrd}}
\end{figure}

\begin{figure*}
	\centering
	
	\includegraphics[width=18cm, height=15cm]{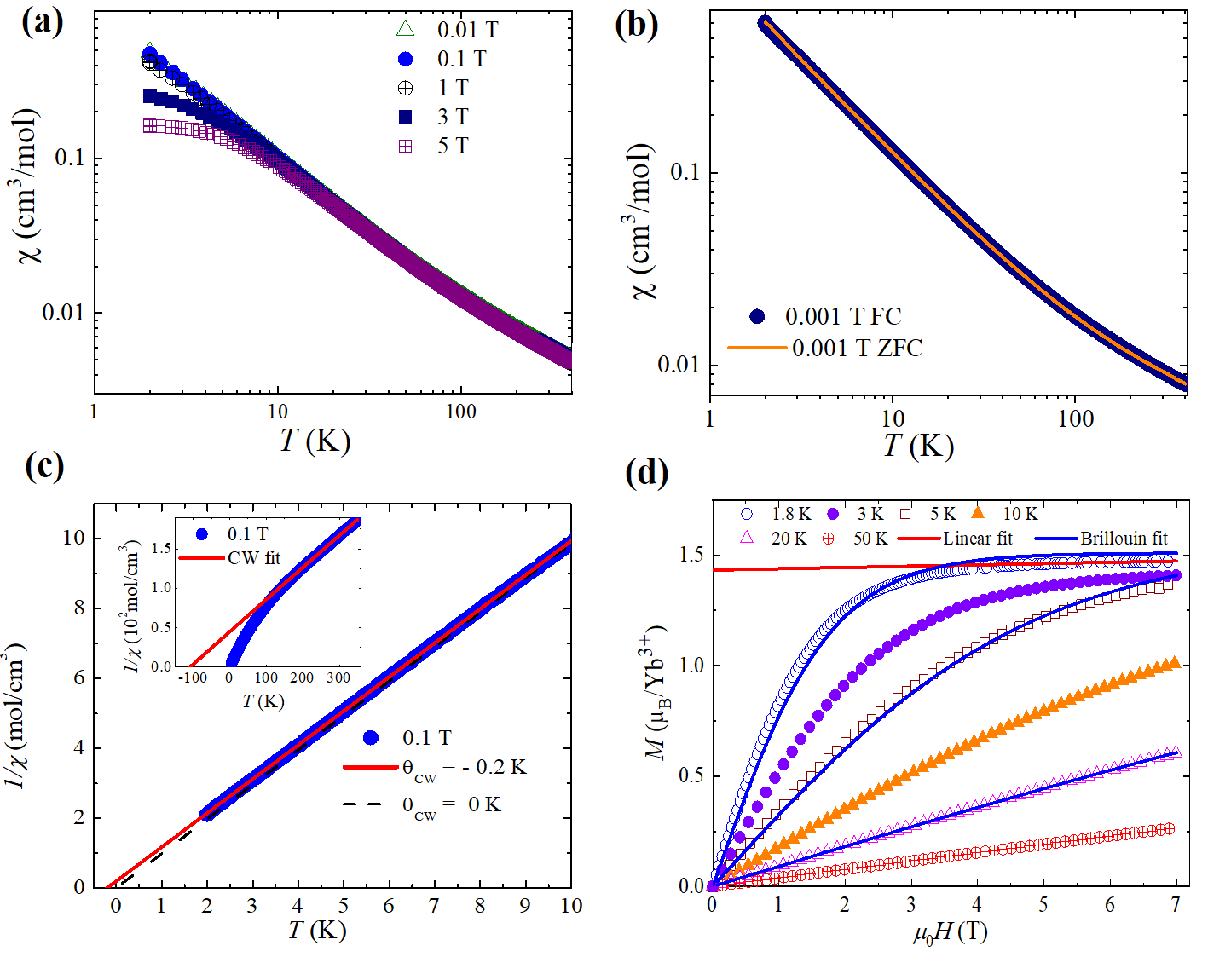}
	\caption{\textbf{(a)} The temperature dependence of magnetic susceptibility in different applied magnetic fields  reordered in ZFC mode. \textbf{(b)} The temperature dependence of zero field cooled and field cooled magnetization data taken in 0.001 T.  \textbf{(c)} Curie-Weiss fit to inverse susceptibility data taken in 0.1 T. Blue filled circle denotes experimental data, whereas solid and dashed lines denote Curie-Weiss fits. Inset shows the same fit at higher temperature range 200 K $\leq T \leq$ 350 K. \textbf{(d)} Magnetic isotherms at different temperatures. Magnetization taken in ramping up and down modes remain identical that add further  credence to the absence  of spin glass physics. From the slope of linear fit (solid red line) to high field data taken at 1.8 K a Van Vleck contribution to susceptibility was extracted.   Blue lines are  Brillouin fits  for \jeff/ $=1/2$ at different temperatures.}
	\label{fig:chiall}
\end{figure*}

\subsection{Crystal structure}
For the Rietveld refinement of \ybsc/ hyperkagome material, initial crystallographic parameters were taken from its iso-structural compound Gd$_3$Sc$_2$Ga$_3$O$_{12}$ which crystallizes in the cubic space group $Ia\bar{3}d$ \cite{Mukherjee_2017}. Rietveld refinement was performed using FullProf Suite on powder XRD data recorded at room temperature and the refined parameters are $R_\mathrm{p} = 8.3 \%$, $R_\mathrm{wp} = 9.2 \%$ and $\chi^2=5.4$ \cite{RODRIGUEZCARVAJAL199355} (see Fig.~\ref{fig:xrd}).  Refined atomic coordinates are presented in Table ~\ref{tab1}. The lattice constant $a$ turned out to be 12.391 {\r{A}}, which is smaller than that found in other  rare-earth-based iso-structural compounds (R$_3$Sc$_2$Ga$_3$O$_{12}$, R = Gd, Tb, Dy and  Ho) \cite{Mukherjee_2017}. This reduction is due to the small ionic radius of Yb$^{3+}$ ion compared to other rare-earth ions.  A tiny impurity peak at around $2\theta \approx 31^\circ$ indicates  a minor (1.5\%) phase of unreacted  Sc$_2$O$_3$ which is non-magnetic and its impact on the overall magnetic behavior of the oxide is negligible. The crystal structure was drawn by using VESTA, which is shown in Fig.~\ref{structure}a \cite{Momma:db5098}. The two inter-penetrating hyperkagome network build up of \yb3+/ ions is shown  in  Fig.~\ref{structure}b. Each Yb atom is surrounded by eight oxygen atoms forming a YbO$_8$ polyhedron. Four of them are at a distance  2.4 {\r{A}}  and the rest are at 2.455 {\r{A}} forming a dodecahedron structure. The nearest-neighbor  Yb$-$Yb bond length is 3.794  {\r{A}}, whereas the second nearest-neighbor bond is 5.796 {\r{A}}. There are two Yb$-$O$-$Yb bridging exchange paths between two nearest-neighbor Yb atoms both making an angle  102.8\textdegree. The presence of corner-shared triangles might induce  magnetic frustration in this material.  

\begin{table}[!htbp]
	\renewcommand{\arraystretch}{1.5}
	\renewcommand{\tabcolsep}{0.3cm}
	\caption{Structural parameters of \ybsc/     determined from the Rietveld refinement of  powder x-ray difraction pattern taken at room temperature using Fullprof software. Space group $Ia\bar{3}d$ and cell parameters  $a = b = c=$ 12.391 {\r{A}}, $\alpha$ = $\beta$ =  $\gamma$ = 90$^\circ$, $\chi^2 =5.4$.}{\label{tab1}}
	\begin{tabular}{c c c c c c} 
		\hline
		\hline
		Atoms & Wyckoff & x & y & z &Occ. \\
		\hline
		Ga&24\textit{d}   & 0&0.25   &0.375&1	
		\\
		Sc&16\textit{a}  &0   &0   &0  &1	
		\\
		Yb&24\textit{c} &0 &0.25   &0.125    &1
		
		\\
		O&96\textit{h} &-0.0316  &0.0561 &0.1508 &1
		
		\\
		
		\hline
		\hline
	\end{tabular}
\end{table}

\subsection{Magnetization}
Magnetic susceptibility measurements  were carried out in the temperature range 1.8 K $\leq T\leq$ 350 K
under several applied magnetic fields. Fig.~\ref{fig:chiall}a represents the temperature dependence of magnetic susceptibility data taken in applied magnetic fields ranging from 0.01 T $\leq T \leq $  5 T. The absence of an anomaly in magnetic susceptibility data indicates that there is no magnetic phase
transition down to 1.8 K.  The absence of any splitting in zero field cooled (ZFC) and field cooled (FC) data taken in 0.001 T rules out spin freezing  down to 1.8 K (see Fig.~\ref{fig:chiall}b). This is further confirmed by the  absence of hysteresis in the magnetization isotherm taken at 1.8 K (see Supplementary Fig.1b). 
High temperature inverse susceptibility data were well reproduced using the  Curie-Weiss formula, $\chi(T)=\frac{C}{T-\theta_\mathrm{CW}}$ in the temperature range 200 K $\leq T\leq$ 350 K (see Fig.~\ref{fig:chiall}c).  The fit yields a  negative $\theta_\mathrm{CW}\approx - 114 \pm 1$ K,  which is attributed to crystal electric field effects and a change in the population of crystal electric field levels can induce a strong  curvature in $\chi^{-1}(T)$ curve (here below 150 K).  The Curie-Weiss constant $C$ turns out to be 2.49 cm$^3\cdot$K/mol resulting 	in effective moment  $\mu_{\mathrm{eff}} = 4.5$ $\mu_\mathrm{B}$, close to  the free ion effective moment (4.54 $\mu_\mathrm{B}$) of Yb$^{3+}$ (4$f^{13}$, $^2$F$_{7/2}$; $J$ = 7/2) ion  ($g\sqrt{J(J+1)}\mu_\mathrm{B}$, $g-$Land\'e factor). In the presence of crystal electric field (CEF), the eight-fold degenerate ground state of Yb$^{3+}$ ion splits into four Kramers doublets \cite{PhysRev.159.245}. In this scenario, Curie-Weiss temperature obtained from high temperature fit  is  inadequate to describe the  magnetic exchange interaction as some of the Kramers doublets with higher energy might be populated and Curie-Weiss temperature is dominated by these crystal field levels (see Supplementary text).  The deviation of the temperature dependence of inverse magnetic susceptibility around 150 K indicates the presence of another energy scale of interaction between Yb$^{3+}$ moments at low temperature. In principle, the correlation between 4$f$ moments develop at very low temperature  and this correlations accompanied by competing degrees of freedom account for the ground state properties in rare-earth-based frustrated quantum magnets \cite{Arh2022}. In order to get an idea regarding  the exchange interaction between Yb$^{3+}$ moments, we have fitted the temperature dependence of inverse magnetic  susceptibility with Curie-Weiss formula in the temperature range $1.8 \leq T \leq 10$ K (see Fig. \ref{fig:chiall}c). A weak Curie-Weiss temperature of $-0.2(1)$ K  and  an effective moment of $2.9(3)$ $\mu_\mathrm{B}$ were obtained from the best fit (see Supplementary Fig. 1a). The negative and small Curie-Weiss temperature indicates the presence of weak antiferromagnetic  interaction between the \yb3+/ spins. Small Curie-Weiss temperature is usually found in rare-earth  magnets \cite{PhysRevB.104.064425,PhysRevLett.100.237204,Arh2022}.  From the value of the effective moment, the $g$ factor turns out to be $3.3(3)$.  The reduced effective moment compared to that expected for Yb$^{3+}$ free ion is attributed  to the \jeff/ $=1/2$ Kramers doublet ground state at low temperatures \cite{Shen2016,PhysRevLett.100.237204,PhysRevB.104.024427,Arh2022}. 
\begin{table}[!htbp]
	\renewcommand{\arraystretch}{1.5}
	\renewcommand{\tabcolsep}{0.08cm}
	\caption{Comparison of lattice parameter, Curie-Weiss temperature ($\theta_\mathrm{CW}$),  ordering temperature of scandium garnet R$_3$Sc$_2$Ga$_3$O$_{12}$. }{\label{tab12}}
	\begin{tabular}{c c c c c c c} 
		\hline
		\hline
		R  & Gd & Tb & Dy & Ho &Yb \\
		\hline
		$a$ ({\r{A}}) &12.573   & 12.539 &12.502  &12.475&12.391\\
		\hline
		$\theta_\mathrm{CW}$ (K)& $−2.2$& $-1.2$&$-0.8$& $-2.9$& $-0.2(1)$\\
	
		\hline
		$T_\mathrm{N}$(K)&$<$0.4 &0.7&1.11&2.4& $<$0.13 \\
		\hline
		Reference &\cite{Mukherjee_2017}   & \cite{Mukherjee_2017} &\cite{Mukherjee_2017} &\cite{Mukherjee_2017}&\textbf{This work}\\				
	\end{tabular}
\end{table}   Fig.~\ref{fig:chiall}d depicts the magnetization isotherms at different temperatures. In higher field and low temperature, magnetization increases linearly due to the Van-Vleck paramagnetism. From the linear behavior of the curve, we have determined a Van-Vleck term of $\chi_{vv}=0.0059 $ $\mu_{\mathrm{B}}/\mathrm{T}= 0.003$ cm$^3$/mol. The saturation magnetic moment was found to be 1.43 $\mu_{\mathrm{B}}/$Yb$^{3+}$ after subtracting  the Van Vleck term, which is  consistent with $M_\mathrm{s}=g_\mathrm{avg}J_\mathrm{eff}(1/2) \mu_\mathrm{B}$. The fit of the $M(H)$ data taken at 1.8 K with Brillouin function also yields a $g$ value of $\approx 3.0(2)$ close to the value obtained from Curie-Weiss fit to low-temperature inverse magnetic susceptibility data (see Fig.~\ref{fig:chiall}d). However, the Brillouin fit deviates slightly, probably indicating the presence of a weak antiferromagnetic interactions between Yb$^{3+}$ moments at low temperature. The data taken at 20 K are well fitted with the Brillouin function as the material is in the paramagnetic region. In these fittings, the $J$ value was fixed to $1/2$ to account for the \jeff/ $= 1/2$ ground state. The dipolar interaction energy and dipolar Curie-Weiss temperature were deduced from the formula $E_\mathrm{d}=\left(\frac{8}{\sqrt{6}}\right)^3\frac{\left(g\mu_{\mathrm{B}}S\right)^2}{a^3}$ and $\theta_\mathrm{CW}^{\mathrm{dip}}=32\pi \frac{\left(g\mu_{\mathrm{B}}\right)^2 S(S+1)}{3a^3k_\mathrm{B}}$ which was used for the isostructural compound Yb$_3$Ga$_5$O$_{12}$ \cite{PhysRevB.104.024427}. Here $a$ is the lattice parameter. $E_\mathrm{d}$ and $\theta_\mathrm{CW}^{\mathrm{dip}}$ are turned out to be 31 mK and 90 mK, respectively. Different magnetic interactions contributes additively to the Curie-Weiss temperature (i.e, $\theta_\mathrm{CW}=\theta_\mathrm{CW}^{\mathrm{dip}}+\theta_\mathrm{CW}^{\mathrm{ex}}$) \cite{jense1991rare,Sosin2022}, the Curie-Weiss temperature due to exchange interaction (i.e, $\theta_\mathrm{CW}^{\mathrm{ex}}$) turned out to be $\sim -0.3(1)$ K. The exchange interaction $J_\mathrm{ex}$   could be obtained using the formula $J_\mathrm{ex}=\frac{3k_\mathrm{B}\theta_\mathrm{CW}^{\mathrm{ex}}}{zS(S+1)}$  \cite{Sackville_Hamilton_2014}, where $z$ represents the number of nearest-neighbors. Using $z= 4$  for hyperkagome structure and effective spin $S\equiv J_\mathrm{eff} =1/2$, we found  $J_\mathrm{ex}=-0.3(1)$ K. The presence of weak exchange interaction is usual in rare-earth oxides which is attributed to the fact that the $4f$ orbitals of rare-earth ions are localized and well shielded yielding weak overlap between two nearest-neighbor orbitals.

\subsection{ Electron Spin Resonance ( ESR)}
\begin{figure}[!htb]
	\centering
	\includegraphics[height=8cm,width=8cm]{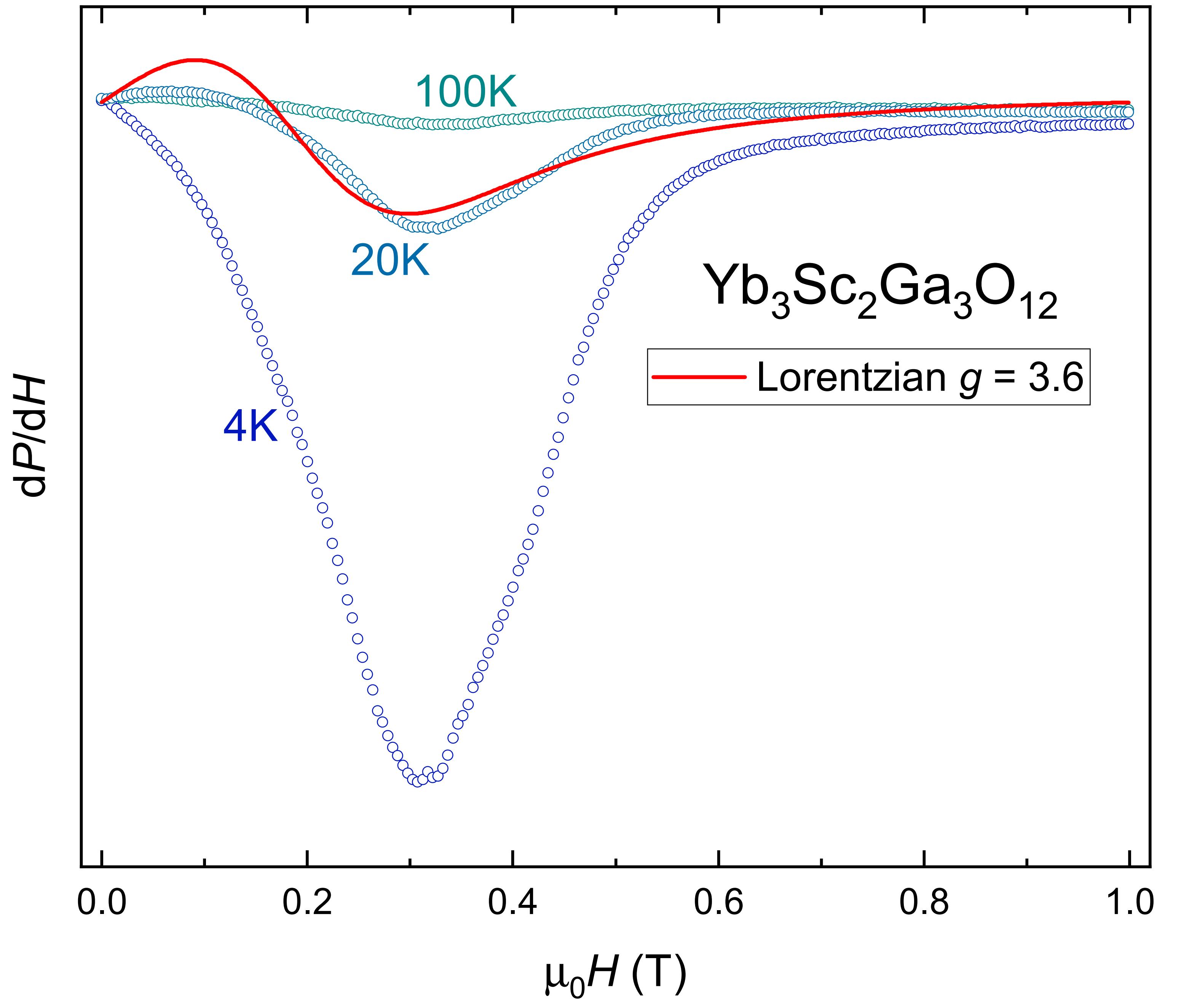}
	\caption{ESR spectra of a polycrystalline sample of \ybsc/ at indicated temperatures and X-band frequency. $dP/dH$ denotes the first field derivative of the absorbed microwave power. Line denotes fitting of the 20 K data with a Lorentzian shape with a corresponding $g$-value of 3.6.}
	\label{fig:esr}
\end{figure}

Electron spin resonance (ESR) is an excellent low energy, local probe to determine the effective $g$–factor as well as characterizing single–ion and exchange anisotropies, which are crucial to establish realistic microscopic Hamiltonian of frustrated quantum materials. It was shown that ESR can shed microscopic insights into the QSL state and associated exotic excitations in frustrated magnets with significant spin-orbit interaction \cite{PhysRevLett.120.037204}. Fig.~\ref{fig:esr} depicts ESR signals of a polycrystalline sample of \ybsc/ at representative temperatures. Due to the rather large linewidth, Yb containing secondary phases may be excluded as origin of the signal. We described the line by a Lorentzian shape including also the negative resonance field component due to the influence of the counter-rotating component of the linearly polarized microwave field. Such a fit yields for the 20 K data a resonance field of 0.18 T (which corresponds to a $g$-value of 3.6) and a linewidth of 0.2 T. The large line width as compared to the resonance field pose a strong constraint to precisely quantify temperature dependencies of ESR parameters in the measured temperature range $4 \leq T \leq 175$ K.
Nevertheless, a clear tendency of line broadening towards low and high temperatures as well as a down shift of the resonance field towards low temperature can be identified. This behavior may be due to a relaxation via the first excited crystal electric field level of the Yb$^{3+}$ ion and the slowing down of Yb$^{3+}$ spin fluctuations in the low temperature regime \cite{PhysRevB.107.064421}. The strong deviation from a symmetric Lorentzian line shape points to off-diagonal terms in the dynamic susceptibility. This is typical for systems with anisotropic spin lattices and becomes even more pronounced when very broad lines are present \cite{H_Benner_1983}. A detailed discussion in this direction would be reasonable only if ESR data of single crystals would be available.

\begin{figure*}[ht]
	\centering
	\includegraphics[height=15cm,width=18cm]{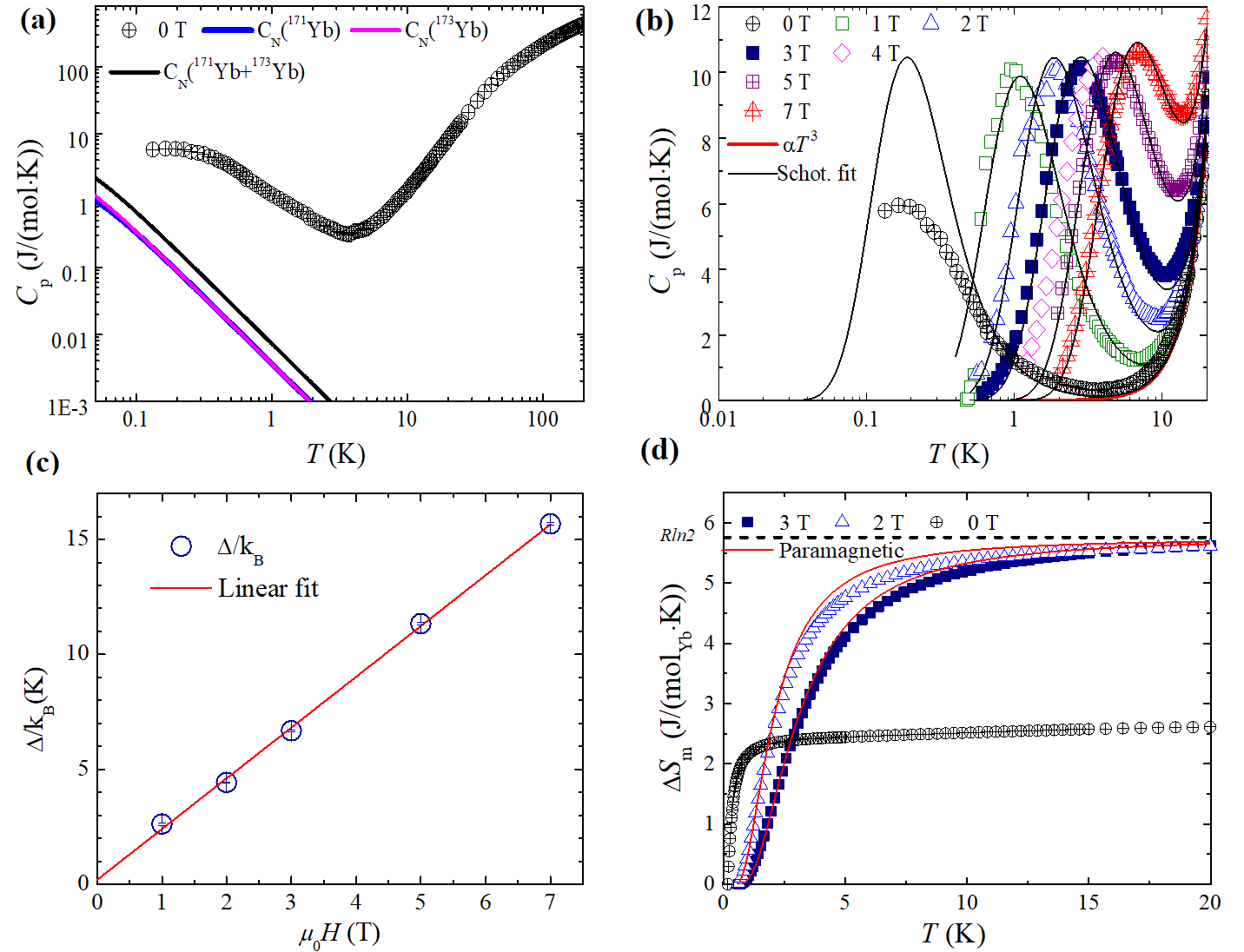}
	
	\caption{\textbf{(a)} Temperature dependence of specific heat ($130$ mK $\leq T \leq 200$ K) for \ybsc/ in zero field. Nuclear Schottky contribution to the specific heat for 3 $\times$ 0.1431 mol of $^{171}$Yb and   3 $\times$ 0.1613 mol of $^{173}$Yb are denoted by blue and pink  line, respectively. \textbf{(b)} Low temperature specific heat ($130$ mK $\leq T \leq 20$ K) in several applied magnetic fields. Corresponding Schottky fits are denoted by black solid lines. Phonon contributions were taken into account while fitting by adding $\alpha T^3$ term  indicated by red solid line. \textbf{(c)} Linear behavior of Schottky gap ($\Delta/k_\mathrm{B}$) with applied magnetic field and the solid red line is a linear fit. From the slope of the fit we found $g=3.3(1)$ and the intercept turns out to be $0.2(1)$ K. \textbf{(d)} Magnetic entropy in  applied magnetic fields. Corresponding paramagnetic entropy in applied fields are denoted by red solid lines. Experimental entropy is slightly smaller than the ideal paramagnet which suggests the existence  of small exchange interaction and/or magnetic frustration. For higher fields, it saturates at 97 \% of the full entropy ($R \ln 2$ ) expected for magnetic materials with \jeff/ $=1/2$ degrees of freedom  consistent with magnetization results. }
	\label{fig:cpschottky}
\end{figure*}
\subsection{Specific heat}

Specific heat provides an excellent probe to gain insights into the ground state properties of this rare-earth hyperkgome lattice. Fig.~\ref{fig:cpschottky}a shows the specific heat  as a function of  temperature down to 130 mK in zero field. The absence of any anomaly in zero field, thus confirms the evasion of long-range magnetic ordering down to at least 130 mK despite a weak antiferromagnetic exchange interaction between \jeff/ $=1/2$ moments. The peak in specific heat due to the nuclear Schottky effect usually occurs below 10$^{-2}$ K \cite{tari2003specific,gopal2012specific,PhysRevB.106.014409}. To calculate the expected nuclear specific heat contribution, we have considered the fact that the natural abundance of  $^{171}$Yb ($I=1/2$) and $^{173}$Yb ($I=5/2$) are 14.31\% and 16.13\%, respectively (see Fig.~\ref{fig:cpschottky}a) \cite{JFilippi1980,PhysRev.128.1136,doi:10.1143/JPSJ.77.104710}. The energy separation between two consecutive levels for  $^{171}$Yb and $^{173}$Yb, are 63.9 mK and 17.55 mK, respectively, which are taken from that reported for the isostructural compound Yb$_3$Ga$_5$O$_{12}$ \cite{JFilippi1980}.  The quadrupole coupling constant $P$ (see supplementary equation 3) is zero for $^{171}$Yb and non-zero for $^{173}$Yb. However, we have neglected the quadrupole term in our calculation ( see SI for details) due to its insignificant impact on shifting the Schottky peak.  Upon the application of
an external magnetic field, a broad maximum appears at low
temperature in specific heat data of \ybsc/, that shifts significantly to higher
temperature upon increasing magnetic field (see Fig.~\ref{fig:cpschottky}b),
which is associated with the electronic Schottky anomaly originating
from the Zeeman splitting of the lowest lying energy levels. To evaluate Schottky contribution, field dependent $ C_\mathrm{p}(H)$ data were fitted with the equation $C_\mathrm{p}=	C_\mathrm{L}+C_\mathrm{S}(\Delta)$ (see Fig.~\ref{fig:cpschottky}c), where $C_\mathrm{S}(\Delta)$ is the specific heat with   energy level splitting $\Delta$  due to an applied magnetic field  $\mu_\mathrm{0}H$ \cite{PhysRevLett.125.267202,PhysRevB.86.140405}. $	C_\mathrm{L}$ is the lattice contribution to the specific heat.  We have fitted the low-temperature zero-field specific heat data in the temperature range 4 K $\leq T \leq$ 22 K with the equation  relevant for lattice specific heat in the simplest approximation
\begin{equation}
	C_\mathrm{L}=\alpha T^3
\end{equation}
and then extrapolated it to lower temperature. The Schottky specific heat $C_\mathrm{S}(\Delta)$ is given by

\begin{equation}
	C_{S}(\Delta)= 3fR \left(\frac{\Delta}{k_\mathrm{B}T}\right)^2 \frac{\mathrm{exp}({\frac{\Delta}{k_\mathrm{B}T})}}{\left[1+ \mathrm{exp}{(\frac{\Delta}{k_\mathrm{B}T})}\right]^2}.
\end{equation}
\begin{figure*}[!ht]
	\centering
	\includegraphics[width=18cm, height=5.5cm]{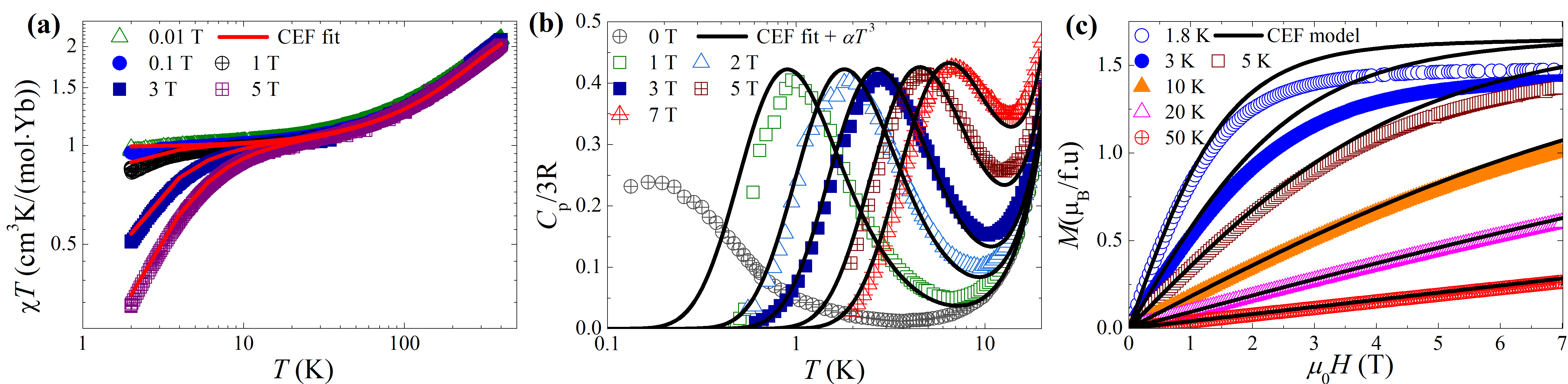}
	\caption{\textbf{(a)} The temperature dependence of $\chi T$ emphasizing the $\chi$ behavior at low temperatures. \textbf{(b)} The temperature dependence of specific heat measured in different magnetic fields. \textbf{(c)} Magnetization as a function of external magnetic field at several temperatures. Lines correspond to the CEF model presented in the text.}
	\label{fig-CEF-all} 
\end{figure*}

The factor 3 indicates that there are 3 Yb ions in a formula unit. Here, $f$ denotes the fraction of \yb3+/ spins participating in the splitting of the ground state.  This value is close to one ($f>0.9$ for fields $\mu_\mathrm{0}H>3$ T)  indicating that  almost all the spins are  contributing to the specific heat. It is worth mentioning that, for transition metal oxides, $f$ denotes the fraction of orphan or defect spins (5-10 \%) that are present in the system owing to disorder \cite{PhysRevLett.100.157205,PhysRevLett.106.147204}. It is found that the value of $\Delta$ increases with an external applied magnetic field typical for many frustrated magnetic materials \cite{PhysRevB.106.104404, PhysRevB.106.014409}. The Land\'e-$g$ factor turns out to be 3.3(1) from the slope of the linear fit to $\Delta/k_\mathrm{B}$ vs. $\mu_\mathrm{0}H$ plot (see inset of  Fig.~\ref{fig:cpschottky}c),  which is consistent with the result  obtained from magnetization data. We get a small zero-field splitting of 0.2(1) K from the intercept of the extrapolated linear fit. However, the deviation of the zero field data from Schottky specific heat indicates that the peak around 200 mK might be due to the presence of short-range spin correlation between  Yb$^{3+}$ ions in this 3D spin-lattice. It is worthwhile to mention that a crystal electric field can also split the energy levels and give rise to Schottky type anomaly. However, the energy difference from the ground state to the first excited state in most of the rare-earth materials is of the order of 5 meV or more, which corresponds to a temperature of 58 K \cite{PhysRevB.100.184415,Paddison2017,PhysRevB.98.220409,Arh2022}.  Magnetic specific heat $C_\mathrm{m}$ was obtained after subtracting the lattice ($C_\mathrm{L}$) and nuclear contribution [$C_\mathrm{N}(^{171}$Yb + $^{173}$Yb)]  to specific heat  from the raw specific heat data $C_\mathrm{p}$ in the entire temperature range. \begin{table*}[!htb]
	\caption{CEF parameters (in meV) for CEF Hamiltonian of Eq.\,(\ref{CEF-Ham}) and $g$ factors of the ground state for the CEF model.
		\label{CEF-par}}
	\begin{ruledtabular}
		\begin{tabular}{c|c|c|c|c|c|c|c|c|c|c|c}
			$B_2^0$ & $B_2^2$ & $B_4^0$ & $B_4^2$ & $B_4^4$ & $B_6^0$ & $B_6^2$ & $B_6^4$ & $B_6^6$ & $g_x$ & $g_y$ & $g_z$ \\
			\hline
			$-$0.06229 & 0.3036 & 0.08183 & $-$0.04343 & $-$0.03914 & 0.0000166 & $-$0.01486 & 0.02468 & $-$0.009285 & 2.93 & 2.25 & 4.24
		\end{tabular}
	\end{ruledtabular}
\end{table*}However, we did not subtract the nuclear contribution for higher field data as our lowest accessible temperature in applied field is 0.5 K, where the nuclear contribution is almost negligible [i.e., 0.026 J/(mol$\cdot$K)]. The magnetic entropy can shed insights into the ground state properties of frustrated magnetic materials.  We have extracted the entropy by integrating $C_\mathrm{m}/T$ from the lowest accessible temperature (130 mK for zero field) to 20 K and the results are shown in Fig.~\ref{fig:cpschottky}d. Due to experimental limitations in accessing temperature below 130 mK,  the zero field specific heat is not fully recovered yielding a lower value of entropy  than that expected for  spin half magnetic materials. The zero-field entropy does not saturate to $R \ln 2$ even up to 20 K. This residual entropy below 130 mK  might indicates the role of magnetic frustration at very low temperature. However, in the application of a magnetic field, the entropy is dominated by the Schottky effect in Kramers doublet systems. The obtained high temperature $\theta_\mathrm{CW}$ ($-114$ K), which indicates crystal field splitting of the ground state, is very large. We expect that the excited state doublets are well separated from the ground state so that we can describe the low-temperature properties by accounting only the lowest $J_\mathrm{eff}=1/2$ ground state. The entropy taken in 3 T saturates with a value 5.6 J/(mol$\cdot$K) which is close to the full entropy  of $R \ln 2$ [ i.e, 5.76 J/(mol$\cdot$K)] for \jeff/ $=1/2$ spin state. This suggests that the ground state of this three-dimensional spin-lattice could be mapped with a low energy  effective \jeff/ $=1/2$ state at low temperature, which is consistent with magnetization results \cite{Yamashita2008,Yamashita2011}.

\subsection{Crystal electric field (CEF) calculations }

We performed a combined fit of specific heat  data measured in  1, 2, 3, 5, and 7 T as well as magnetic susceptibility data measured in 0.01, 0.1, 1, 3, and 5 T.
According to the Hund's rule, the ground state multiplet of the Yb$^{3+}$ ion is $^2F_{7/2}$, which is in a crystal field split into four Kramers doublets composed of $|\pm m_J\rangle$ states [$m_J$\,=\,$(2n-1)/2$, where $n$\,=\,1-4].
The composition of the four Kramers doublets directly depends on the CEF Hamiltonian, which can be written as
\begin{equation}
	H_{\text{CEF}} = \sum_{i,j} B_j^i O_j^i,
	\label{CEF-Ham}
\end{equation}

where $O_j^i$ are Stevens operators \cite{stevens1952matrix} and $B_j^i$ are the corresponding scaling parameters. 
The relevant $B_j^i$ (Table\,\ref{CEF-par}) are determined by point symmetry at the Yb$^{3+}$ site and in general comply with those determined for the isostructural compound Yb$_3$Ga$_5$O$_{12}$ \cite{sandberg2021emergent}.%
\begin{table}[htb]
	\caption{CEF energy levels (in meV) for the CEF model.
		\label{CEF-level}}
	\begin{ruledtabular}
		\begin{tabular}{c|c|c|c}
			$E_0$ & $E_1$ & $E_2$ & $E_3$\\
			\hline
			0 & 69 & 96 & 143 \\
		\end{tabular}
	\end{ruledtabular}
\end{table}
Indeed, we obtained a very good agreement with the data measured in the applied magnetic field [lines in Fig.\,\ref{fig-CEF-all}(a) and (b)] for the derived CEF parameters (Table\,\ref{CEF-par}), when considering additional $T^3$ term for the specific heat, accounting for the phonon contribution at low temperatures. Finally, we have calculated the corresponding magnetization curves $M(H)$  [lines in Fig.\,\ref{fig-CEF-all}(c)].
The agreement of the simulation and experiment is quite reasonable for both experiments.
The derived magnetic anisotropy of the CEF ground state is determined by the $g$ factor, also given in Table\,\ref{CEF-par}, implying a rather uniaxial anisotropy, as $g_z$ is considerably larger than $g_x$ and $g_y$. In fact, the derived $g$ values are in agreement with the average $g = 3.6$ derived  from the ESR data measured at $20$ K.
All energy levels are summarized in table Table\,\ref{CEF-level}, indicating that the ground state is well separated from the excited states, as found also in Yb$_3$Ga$_5$O$_{12}$ \cite{sandberg2021emergent}.
We note that, the broad maximum  in specific heat  around 200 mK in zero field is clearly not related to CEF effects and as suggested above most likely associated with the presence of short-range spin correlations.

\section{Discussion}
The intertwining of spin-orbit coupling and electron correlation in frustrated magnets can stabilize exotic quantum states. In this context, rare-earth-based $4f$  frustrated magnets offer  an  exciting platform. In these materials, anisotropy and crystal electric field  generated by the neighboring ligands govern the ground state properties. The choice of a suitable chemical element has a great role in tuning the crystal electric field, anisotropy and exchange interaction and  hence the ground state properties of rare-earth-based frustrated magnet. The hyperkagome  material \ybsc/ belongs to the garnet family with cubic space group $Ia\bar{3}d$ and lattice constant $a=12.391$ {\r{A}}. Unlike R$_3$Sc$_2$Ga$_3$O$_{12}$ garnets (R = Tb, Dy, Ho), which orders at low temperature, \ybsc/ does not show any long-range magnetic ordering at least down to 130 mK. This might be due to the low magnetic moment of \yb3+/  and low exchange energy compared to other R$_3$Sc$_2$Ga$_3$O$_{12}$ compounds \cite{Mukherjee_2017}. Furthermore, the absence of ZFC-FC splitting rules out spin-freezing in this material. The small negative Curie-Weiss temperature indicates the presence of a weak antiferromagnetic interaction which arises mainly due to super-exchange interaction.  The $\theta_\mathrm{CW}$  value in a three-dimensional hyperkagome lattice Li$_3$Yb$_3$Te$_2$O$_{12}$ with a similar oxygen environment is almost the same order of magnitude at high temperature as well as at low temperature  as compared to the present hyperkgome material \cite{PhysRevB.106.104404}.  ESR spectra taken on polycrystalline samples of \ybsc/ are found to be relatively broad with a highly asymmetric shape which is related to magnetic anisotropy  and most likely also the distribution of Yb$^{3+}$ moments \cite{H_Benner_1983}.  The specific heat experiment which is very sensitive to probe magnetic phase transition rules out a long-range  magnetic order down to 130 mK in this  frustrated magnet.  The broad maximum around 200 mK in specific heat data taken in zero field is ascribed to the development of short-range spin correlations at low temperature. The development of short-range spin correlations suggesting the dominance of quantum effects at low temperature that lead to a dynamic liquid-like state in this 3D frustrated antiferromagnet. Similar broad maximum in specific heat at low temperature was also observed in its sister compound Yb$_3$Ga$_5$O$_{12}$ \cite{JFilippi1980,PhysRevLett.91.167201}. The low temperature magnetic behavior can be described  by \jeff/ $=1/2$ Kramers doublet ground state as the energy levels  of the excited Kramers doublets is well above the ground state, which is confirmed by our CEF calculations. This family of R$_3$Sc$_2$Ga$_3$O$_{12}$ rare-earth  magnetic materials with a suitable choice of rare-earth and non-magnetic cations  has a great potential   to host non-trivial quantum states with exotic excitations.

\section{Conclusion}
In summary, we have successfully synthesized and characterized a three-dimensional hyperkagome garnet,  namely \ybsc/. Magnetization data indicate the absence of long-range magnetic  ordering and spin freezing down to 1.8 K. Specific heat measurements reveal that there is no  long-range magnetic ordering  down to 130 mK. A weak antiferromagnetic exchange interaction is present in this system, which is typical for rare-earth-based magnetic materials due to the localized 4$f$ moments. In addition, the absence of ZFC-FC splitting rules out  the formation of a glassy phase in this antiferromagnet.   Our thermodynamic results indicate  that the \yb3+/ ions are in the Kramers doublet state with an effective \jeff/ $=1/2$ spin degrees of freedom at low temperature. Furthermore, ESR results indicate the existence of magnetic anisotropy. The presence of a broad maximum around 200 mK  in the temperature dependence of specific heat suggests  that  a short-range spin correlation  phenomena at play in this frustrated magnet. Our CEF modeling suggests that the  ground state is well
separated from the excited states and  yields a very good
agreement with thermodynamic and ESR results. Given the fact that the exchange interaction is weak in 4$f$ systems, further low-temperature thermodynamic and muon spin relaxation experiments are required to track the ground state properties unambiguously. Microscopic experiments such as neutron scattering are desired to shed insights into the spin-orbit driven anisotropy driving novel ground state arises from the crystal-field splitting of the rare-earth multiplet. Understanding the spin anisotropy and low energy spin excitation spectra sets an interesting track to  reveal interesting insights into the ground state properties of  this   novel frustrated  magnet. This family of rare-earth hyperkagome R$_3$Sc$_2$Ga$_3$O$_{12}$ (R $=$ rare-earth)  offers a promising platform for the experimental realization of unconventional ground states borne out of spin correlation, frustration and  spin-orbit driven anisotropy.

\section{Acknowledgment}

PK acknowledges the funding by the Science
and Engineering Research Board, and Department of Science
and Technology, India through Research Grants. MP acknowledges the funding by the Slovenian Research Agency (project J2-2513, and program No. P1-0125).

\bibliography{references}

\begin{thebibliography}{76}%
\makeatletter
\providecommand \@ifxundefined [1]{%
 \@ifx{#1\undefined}
}%
\providecommand \@ifnum [1]{%
 \ifnum #1\expandafter \@firstoftwo
 \else \expandafter \@secondoftwo
 \fi
}%
\providecommand \@ifx [1]{%
 \ifx #1\expandafter \@firstoftwo
 \else \expandafter \@secondoftwo
 \fi
}%
\providecommand \natexlab [1]{#1}%
\providecommand \enquote  [1]{``#1''}%
\providecommand \bibnamefont  [1]{#1}%
\providecommand \bibfnamefont [1]{#1}%
\providecommand \citenamefont [1]{#1}%
\providecommand \href@noop [0]{\@secondoftwo}%
\providecommand \href [0]{\begingroup \@sanitize@url \@href}%
\providecommand \@href[1]{\@@startlink{#1}\@@href}%
\providecommand \@@href[1]{\endgroup#1\@@endlink}%
\providecommand \@sanitize@url [0]{\catcode `\\12\catcode `\$12\catcode
  `\&12\catcode `\#12\catcode `\^12\catcode `\_12\catcode `\%12\relax}%
\providecommand \@@startlink[1]{}%
\providecommand \@@endlink[0]{}%
\providecommand \url  [0]{\begingroup\@sanitize@url \@url }%
\providecommand \@url [1]{\endgroup\@href {#1}{\urlprefix }}%
\providecommand \urlprefix  [0]{URL }%
\providecommand \Eprint [0]{\href }%
\providecommand \doibase [0]{https://doi.org/}%
\providecommand \selectlanguage [0]{\@gobble}%
\providecommand \bibinfo  [0]{\@secondoftwo}%
\providecommand \bibfield  [0]{\@secondoftwo}%
\providecommand \translation [1]{[#1]}%
\providecommand \BibitemOpen [0]{}%
\providecommand \bibitemStop [0]{}%
\providecommand \bibitemNoStop [0]{.\EOS\space}%
\providecommand \EOS [0]{\spacefactor3000\relax}%
\providecommand \BibitemShut  [1]{\csname bibitem#1\endcsname}%
\let\auto@bib@innerbib\@empty
\bibitem [{\citenamefont {Diep}(2005)}]{doi:10.1142/5697}%
  \BibitemOpen
  \bibfield  {author} {\bibinfo {author} {\bibfnamefont {H.~T.}\ \bibnamefont
  {Diep}},\ }\href {https://doi.org/10.1142/5697} {\emph {\bibinfo {title}
  {Frustrated Spin Systems}}}\ (\bibinfo  {publisher} {WORLD SCIENTIFIC},\
  \bibinfo {year} {2005})\BibitemShut {NoStop}%
\bibitem [{\citenamefont {Lacroix}\ \emph {et~al.}(2011)\citenamefont
  {Lacroix}, \citenamefont {Mendels},\ and\ \citenamefont
  {Mila}}]{10.1007/978-3-642-10589-0}%
  \BibitemOpen
  \bibinfo {editor} {\bibfnamefont {C.}~\bibnamefont {Lacroix}}, \bibinfo
  {editor} {\bibfnamefont {P.}~\bibnamefont {Mendels}},\ and\ \bibinfo {editor}
  {\bibfnamefont {F.}~\bibnamefont {Mila}},\ eds.,\ \href
  {https://doi.org/10.1007/978-3-642-10589-0} {\emph {\bibinfo {title}
  {{Introduction to Frustrated Magnetism: Materials, Experiments, Theory
  (Springer Series in Solid-State Sciences)}}}},\ \bibinfo {edition} {2011th}\
  ed.\ (\bibinfo  {publisher} {Springer},\ \bibinfo {year} {2011})\BibitemShut
  {NoStop}%
\bibitem [{\citenamefont {Savary}\ and\ \citenamefont
  {Balents}(2016)}]{Savary_2016}%
  \BibitemOpen
  \bibfield  {author} {\bibinfo {author} {\bibfnamefont {L.}~\bibnamefont
  {Savary}}\ and\ \bibinfo {author} {\bibfnamefont {L.}~\bibnamefont
  {Balents}},\ }\bibfield  {title} {\bibinfo {title} {Quantum spin liquids: a
  review},\ }\href {https://doi.org/10.1088/0034-4885/80/1/016502} {\bibfield
  {journal} {\bibinfo  {journal} {Reports on Progress in Physics}\ }\textbf
  {\bibinfo {volume} {80}},\ \bibinfo {pages} {016502} (\bibinfo {year}
  {2016})}\BibitemShut {NoStop}%
\bibitem [{\citenamefont {Zhou}\ \emph {et~al.}(2017)\citenamefont {Zhou},
  \citenamefont {Kanoda},\ and\ \citenamefont {Ng}}]{RevModPhys.89.025003}%
  \BibitemOpen
  \bibfield  {author} {\bibinfo {author} {\bibfnamefont {Y.}~\bibnamefont
  {Zhou}}, \bibinfo {author} {\bibfnamefont {K.}~\bibnamefont {Kanoda}},\ and\
  \bibinfo {author} {\bibfnamefont {T.-K.}\ \bibnamefont {Ng}},\ }\bibfield
  {title} {\bibinfo {title} {Quantum spin liquid states},\ }\href
  {https://doi.org/10.1103/RevModPhys.89.025003} {\bibfield  {journal}
  {\bibinfo  {journal} {Rev. Mod. Phys.}\ }\textbf {\bibinfo {volume} {89}},\
  \bibinfo {pages} {025003} (\bibinfo {year} {2017})}\BibitemShut {NoStop}%
\bibitem [{\citenamefont {Anderson}(1973)}]{ANDERSON1973153}%
  \BibitemOpen
  \bibfield  {author} {\bibinfo {author} {\bibfnamefont {P.}~\bibnamefont
  {Anderson}},\ }\bibfield  {title} {\bibinfo {title} {Resonating valence
  bonds: A new kind of insulator?},\ }\href
  {https://doi.org/https://doi.org/10.1016/0025-5408(73)90167-0} {\bibfield
  {journal} {\bibinfo  {journal} {Materials Research Bulletin}\ }\textbf
  {\bibinfo {volume} {8}},\ \bibinfo {pages} {153 } (\bibinfo {year}
  {1973})}\BibitemShut {NoStop}%
\bibitem [{\citenamefont {Balents}(2010)}]{Balents2010}%
  \BibitemOpen
  \bibfield  {author} {\bibinfo {author} {\bibfnamefont {L.}~\bibnamefont
  {Balents}},\ }\bibfield  {title} {\bibinfo {title} {Spin liquids in
  frustrated magnets},\ }\href {https://doi.org/10.1038/nature08917} {\bibfield
   {journal} {\bibinfo  {journal} {Nature}\ }\textbf {\bibinfo {volume}
  {464}},\ \bibinfo {pages} {199} (\bibinfo {year} {2010})}\BibitemShut
  {NoStop}%
\bibitem [{\citenamefont {Do}\ \emph {et~al.}(2017)\citenamefont {Do},
  \citenamefont {Park}, \citenamefont {Yoshitake}, \citenamefont {Nasu},
  \citenamefont {Motome}, \citenamefont {Kwon}, \citenamefont {Adroja},
  \citenamefont {Voneshen}, \citenamefont {Kim}, \citenamefont {Jang},
  \citenamefont {Park}, \citenamefont {Choi},\ and\ \citenamefont
  {Ji}}]{Do2017}%
  \BibitemOpen
  \bibfield  {author} {\bibinfo {author} {\bibfnamefont {S.-H.}\ \bibnamefont
  {Do}}, \bibinfo {author} {\bibfnamefont {S.-Y.}\ \bibnamefont {Park}},
  \bibinfo {author} {\bibfnamefont {J.}~\bibnamefont {Yoshitake}}, \bibinfo
  {author} {\bibfnamefont {J.}~\bibnamefont {Nasu}}, \bibinfo {author}
  {\bibfnamefont {Y.}~\bibnamefont {Motome}}, \bibinfo {author} {\bibfnamefont
  {Y.}~\bibnamefont {Kwon}}, \bibinfo {author} {\bibfnamefont {D.~T.}\
  \bibnamefont {Adroja}}, \bibinfo {author} {\bibfnamefont {D.~J.}\
  \bibnamefont {Voneshen}}, \bibinfo {author} {\bibfnamefont {K.}~\bibnamefont
  {Kim}}, \bibinfo {author} {\bibfnamefont {T.-H.}\ \bibnamefont {Jang}},
  \bibinfo {author} {\bibfnamefont {J.-H.}\ \bibnamefont {Park}}, \bibinfo
  {author} {\bibfnamefont {K.-Y.}\ \bibnamefont {Choi}},\ and\ \bibinfo
  {author} {\bibfnamefont {S.}~\bibnamefont {Ji}},\ }\bibfield  {title}
  {\bibinfo {title} {{Majorana fermions in the Kitaev quantum spin system
  $\alpha$-RuCl$_3$}},\ }\href {https://doi.org/10.1038/nphys4264} {\bibfield
  {journal} {\bibinfo  {journal} {Nature Physics}\ }\textbf {\bibinfo {volume}
  {13}},\ \bibinfo {pages} {1079} (\bibinfo {year} {2017})}\BibitemShut
  {NoStop}%
\bibitem [{\citenamefont {Banerjee}\ \emph {et~al.}(2016)\citenamefont
  {Banerjee}, \citenamefont {Bridges}, \citenamefont {Yan}, \citenamefont
  {Aczel}, \citenamefont {Li}, \citenamefont {Stone}, \citenamefont {Granroth},
  \citenamefont {Lumsden}, \citenamefont {Yiu}, \citenamefont {Knolle},
  \citenamefont {Bhattacharjee}, \citenamefont {Kovrizhin}, \citenamefont
  {Moessner}, \citenamefont {Tennant}, \citenamefont {Mandrus},\ and\
  \citenamefont {Nagler}}]{Banerjee2016}%
  \BibitemOpen
  \bibfield  {author} {\bibinfo {author} {\bibfnamefont {A.}~\bibnamefont
  {Banerjee}}, \bibinfo {author} {\bibfnamefont {C.~A.}\ \bibnamefont
  {Bridges}}, \bibinfo {author} {\bibfnamefont {J.-Q.}\ \bibnamefont {Yan}},
  \bibinfo {author} {\bibfnamefont {A.~A.}\ \bibnamefont {Aczel}}, \bibinfo
  {author} {\bibfnamefont {L.}~\bibnamefont {Li}}, \bibinfo {author}
  {\bibfnamefont {M.~B.}\ \bibnamefont {Stone}}, \bibinfo {author}
  {\bibfnamefont {G.~E.}\ \bibnamefont {Granroth}}, \bibinfo {author}
  {\bibfnamefont {M.~D.}\ \bibnamefont {Lumsden}}, \bibinfo {author}
  {\bibfnamefont {Y.}~\bibnamefont {Yiu}}, \bibinfo {author} {\bibfnamefont
  {J.}~\bibnamefont {Knolle}}, \bibinfo {author} {\bibfnamefont
  {S.}~\bibnamefont {Bhattacharjee}}, \bibinfo {author} {\bibfnamefont {D.~L.}\
  \bibnamefont {Kovrizhin}}, \bibinfo {author} {\bibfnamefont {R.}~\bibnamefont
  {Moessner}}, \bibinfo {author} {\bibfnamefont {D.~A.}\ \bibnamefont
  {Tennant}}, \bibinfo {author} {\bibfnamefont {D.~G.}\ \bibnamefont
  {Mandrus}},\ and\ \bibinfo {author} {\bibfnamefont {S.~E.}\ \bibnamefont
  {Nagler}},\ }\bibfield  {title} {\bibinfo {title} {Proximate kitaev quantum
  spin liquid behaviour in a honeycomb magnet},\ }\href
  {https://doi.org/10.1038/nmat4604} {\bibfield  {journal} {\bibinfo  {journal}
  {Nature Materials}\ }\textbf {\bibinfo {volume} {15}},\ \bibinfo {pages}
  {733} (\bibinfo {year} {2016})}\BibitemShut {NoStop}%
\bibitem [{\citenamefont {Broholm}\ \emph {et~al.}(2020)\citenamefont
  {Broholm}, \citenamefont {Cava}, \citenamefont {Kivelson}, \citenamefont
  {Nocera}, \citenamefont {Norman},\ and\ \citenamefont
  {Senthil}}]{doi:10.1126/science.aay0668}%
  \BibitemOpen
  \bibfield  {author} {\bibinfo {author} {\bibfnamefont {C.}~\bibnamefont
  {Broholm}}, \bibinfo {author} {\bibfnamefont {R.~J.}\ \bibnamefont {Cava}},
  \bibinfo {author} {\bibfnamefont {S.~A.}\ \bibnamefont {Kivelson}}, \bibinfo
  {author} {\bibfnamefont {D.~G.}\ \bibnamefont {Nocera}}, \bibinfo {author}
  {\bibfnamefont {M.~R.}\ \bibnamefont {Norman}},\ and\ \bibinfo {author}
  {\bibfnamefont {T.}~\bibnamefont {Senthil}},\ }\bibfield  {title} {\bibinfo
  {title} {Quantum spin liquids},\ }\href
  {https://doi.org/10.1126/science.aay0668} {\bibfield  {journal} {\bibinfo
  {journal} {Science}\ }\textbf {\bibinfo {volume} {367}},\ \bibinfo {pages}
  {eaay0668} (\bibinfo {year} {2020})}\BibitemShut {NoStop}%
\bibitem [{\citenamefont {Shimizu}\ \emph {et~al.}(2003)\citenamefont
  {Shimizu}, \citenamefont {Miyagawa}, \citenamefont {Kanoda}, \citenamefont
  {Maesato},\ and\ \citenamefont {Saito}}]{PhysRevLett.91.107001}%
  \BibitemOpen
  \bibfield  {author} {\bibinfo {author} {\bibfnamefont {Y.}~\bibnamefont
  {Shimizu}}, \bibinfo {author} {\bibfnamefont {K.}~\bibnamefont {Miyagawa}},
  \bibinfo {author} {\bibfnamefont {K.}~\bibnamefont {Kanoda}}, \bibinfo
  {author} {\bibfnamefont {M.}~\bibnamefont {Maesato}},\ and\ \bibinfo {author}
  {\bibfnamefont {G.}~\bibnamefont {Saito}},\ }\bibfield  {title} {\bibinfo
  {title} {Spin liquid state in an organic mott insulator with a triangular
  lattice},\ }\href {https://doi.org/10.1103/PhysRevLett.91.107001} {\bibfield
  {journal} {\bibinfo  {journal} {Phys. Rev. Lett.}\ }\textbf {\bibinfo
  {volume} {91}},\ \bibinfo {pages} {107001} (\bibinfo {year}
  {2003})}\BibitemShut {NoStop}%
\bibitem [{\citenamefont {Itou}\ \emph {et~al.}(2008)\citenamefont {Itou},
  \citenamefont {Oyamada}, \citenamefont {Maegawa}, \citenamefont {Tamura},\
  and\ \citenamefont {Kato}}]{PhysRevB.77.104413}%
  \BibitemOpen
  \bibfield  {author} {\bibinfo {author} {\bibfnamefont {T.}~\bibnamefont
  {Itou}}, \bibinfo {author} {\bibfnamefont {A.}~\bibnamefont {Oyamada}},
  \bibinfo {author} {\bibfnamefont {S.}~\bibnamefont {Maegawa}}, \bibinfo
  {author} {\bibfnamefont {M.}~\bibnamefont {Tamura}},\ and\ \bibinfo {author}
  {\bibfnamefont {R.}~\bibnamefont {Kato}},\ }\bibfield  {title} {\bibinfo
  {title} {{Quantum spin liquid in the spin-$1/2$ triangular antiferromagnet
  $\mathrm{Et}{\mathrm{Me}}_{3}\mathrm{Sb}{[\mathrm{Pd}{(\text{dmit})}_{2}]}_{2}$}},\
  }\href {https://doi.org/10.1103/PhysRevB.77.104413} {\bibfield  {journal}
  {\bibinfo  {journal} {Phys. Rev. B}\ }\textbf {\bibinfo {volume} {77}},\
  \bibinfo {pages} {104413} (\bibinfo {year} {2008})}\BibitemShut {NoStop}%
\bibitem [{\citenamefont {Zhou}\ \emph {et~al.}(2011)\citenamefont {Zhou},
  \citenamefont {Choi}, \citenamefont {Li}, \citenamefont {Balicas},
  \citenamefont {Wiebe}, \citenamefont {Qiu}, \citenamefont {Copley},\ and\
  \citenamefont {Gardner}}]{PhysRevLett.106.147204}%
  \BibitemOpen
  \bibfield  {author} {\bibinfo {author} {\bibfnamefont {H.~D.}\ \bibnamefont
  {Zhou}}, \bibinfo {author} {\bibfnamefont {E.~S.}\ \bibnamefont {Choi}},
  \bibinfo {author} {\bibfnamefont {G.}~\bibnamefont {Li}}, \bibinfo {author}
  {\bibfnamefont {L.}~\bibnamefont {Balicas}}, \bibinfo {author} {\bibfnamefont
  {C.~R.}\ \bibnamefont {Wiebe}}, \bibinfo {author} {\bibfnamefont
  {Y.}~\bibnamefont {Qiu}}, \bibinfo {author} {\bibfnamefont {J.~R.~D.}\
  \bibnamefont {Copley}},\ and\ \bibinfo {author} {\bibfnamefont {J.~S.}\
  \bibnamefont {Gardner}},\ }\bibfield  {title} {\bibinfo {title} {{Spin Liquid
  State in the $S=1/2$ Triangular Lattice
  ${\mathrm{Ba}}_{3}{\mathrm{CuSb}}_{2}{\mathrm{O}}_{9}$}},\ }\href
  {https://doi.org/10.1103/PhysRevLett.106.147204} {\bibfield  {journal}
  {\bibinfo  {journal} {Phys. Rev. Lett.}\ }\textbf {\bibinfo {volume} {106}},\
  \bibinfo {pages} {147204} (\bibinfo {year} {2011})}\BibitemShut {NoStop}%
\bibitem [{\citenamefont {Khuntia}\ \emph
  {et~al.}(2016{\natexlab{a}})\citenamefont {Khuntia}, \citenamefont {Kumar},
  \citenamefont {Mahajan}, \citenamefont {Baenitz},\ and\ \citenamefont
  {Furukawa}}]{PhysRevB.93.140408}%
  \BibitemOpen
  \bibfield  {author} {\bibinfo {author} {\bibfnamefont {P.}~\bibnamefont
  {Khuntia}}, \bibinfo {author} {\bibfnamefont {R.}~\bibnamefont {Kumar}},
  \bibinfo {author} {\bibfnamefont {A.~V.}\ \bibnamefont {Mahajan}}, \bibinfo
  {author} {\bibfnamefont {M.}~\bibnamefont {Baenitz}},\ and\ \bibinfo {author}
  {\bibfnamefont {Y.}~\bibnamefont {Furukawa}},\ }\bibfield  {title} {\bibinfo
  {title} {{Spin liquid state in the disordered triangular lattice
  Sc$_2$Ga$_2$CuO$_7$ revealed by NMR}},\ }\href
  {https://doi.org/10.1103/PhysRevB.93.140408} {\bibfield  {journal} {\bibinfo
  {journal} {Phys. Rev. B}\ }\textbf {\bibinfo {volume} {93}},\ \bibinfo
  {pages} {140408} (\bibinfo {year} {2016}{\natexlab{a}})}\BibitemShut
  {NoStop}%
\bibitem [{\citenamefont {Shores}\ \emph {et~al.}(2005)\citenamefont {Shores},
  \citenamefont {Nytko}, \citenamefont {Bartlett},\ and\ \citenamefont
  {Nocera}}]{doi:10.1021/ja053891p}%
  \BibitemOpen
  \bibfield  {author} {\bibinfo {author} {\bibfnamefont {M.~P.}\ \bibnamefont
  {Shores}}, \bibinfo {author} {\bibfnamefont {E.~A.}\ \bibnamefont {Nytko}},
  \bibinfo {author} {\bibfnamefont {B.~M.}\ \bibnamefont {Bartlett}},\ and\
  \bibinfo {author} {\bibfnamefont {D.~G.}\ \bibnamefont {Nocera}},\ }\bibfield
   {title} {\bibinfo {title} {{A Structurally Perfect S = 1/2 Kagome
  Antiferromagnet}},\ }\href {https://doi.org/10.1021/ja053891p} {\bibfield
  {journal} {\bibinfo  {journal} {Journal of the American Chemical Society}\
  }\textbf {\bibinfo {volume} {127}},\ \bibinfo {pages} {13462} (\bibinfo
  {year} {2005})}\BibitemShut {NoStop}%
\bibitem [{\citenamefont {Olariu}\ \emph {et~al.}(2008)\citenamefont {Olariu},
  \citenamefont {Mendels}, \citenamefont {Bert}, \citenamefont {Duc},
  \citenamefont {Trombe}, \citenamefont {de~Vries},\ and\ \citenamefont
  {Harrison}}]{PhysRevLett.100.087202}%
  \BibitemOpen
  \bibfield  {author} {\bibinfo {author} {\bibfnamefont {A.}~\bibnamefont
  {Olariu}}, \bibinfo {author} {\bibfnamefont {P.}~\bibnamefont {Mendels}},
  \bibinfo {author} {\bibfnamefont {F.}~\bibnamefont {Bert}}, \bibinfo {author}
  {\bibfnamefont {F.}~\bibnamefont {Duc}}, \bibinfo {author} {\bibfnamefont
  {J.~C.}\ \bibnamefont {Trombe}}, \bibinfo {author} {\bibfnamefont {M.~A.}\
  \bibnamefont {de~Vries}},\ and\ \bibinfo {author} {\bibfnamefont
  {A.}~\bibnamefont {Harrison}},\ }\bibfield  {title} {\bibinfo {title}
  {{$^{17}\mathrm{O}$ NMR Study of the Intrinsic Magnetic Susceptibility and
  Spin Dynamics of the Quantum Kagome Antiferromagnet
  ${\mathrm{ZnCu}}_{3}(\mathrm{OH}{)}_{6}{\mathrm{Cl}}_{2}$}},\ }\href
  {https://doi.org/10.1103/PhysRevLett.100.087202} {\bibfield  {journal}
  {\bibinfo  {journal} {Phys. Rev. Lett.}\ }\textbf {\bibinfo {volume} {100}},\
  \bibinfo {pages} {087202} (\bibinfo {year} {2008})}\BibitemShut {NoStop}%
\bibitem [{\citenamefont {Khuntia}\ \emph {et~al.}(2020)\citenamefont
  {Khuntia}, \citenamefont {Velazquez}, \citenamefont {Barth{\'e}lemy},
  \citenamefont {Bert}, \citenamefont {Kermarrec}, \citenamefont {Legros},
  \citenamefont {Bernu}, \citenamefont {Messio}, \citenamefont {Zorko},\ and\
  \citenamefont {Mendels}}]{Khuntia2020}%
  \BibitemOpen
  \bibfield  {author} {\bibinfo {author} {\bibfnamefont {P.}~\bibnamefont
  {Khuntia}}, \bibinfo {author} {\bibfnamefont {M.}~\bibnamefont {Velazquez}},
  \bibinfo {author} {\bibfnamefont {Q.}~\bibnamefont {Barth{\'e}lemy}},
  \bibinfo {author} {\bibfnamefont {F.}~\bibnamefont {Bert}}, \bibinfo {author}
  {\bibfnamefont {E.}~\bibnamefont {Kermarrec}}, \bibinfo {author}
  {\bibfnamefont {A.}~\bibnamefont {Legros}}, \bibinfo {author} {\bibfnamefont
  {B.}~\bibnamefont {Bernu}}, \bibinfo {author} {\bibfnamefont
  {L.}~\bibnamefont {Messio}}, \bibinfo {author} {\bibfnamefont
  {A.}~\bibnamefont {Zorko}},\ and\ \bibinfo {author} {\bibfnamefont
  {P.}~\bibnamefont {Mendels}},\ }\bibfield  {title} {\bibinfo {title}
  {{Gapless ground state in the archetypal quantum kagome antiferromagnet
  ZnCu$_3$(OH)$_6$Cl$_2$}},\ }\href {https://doi.org/10.1038/s41567-020-0792-1}
  {\bibfield  {journal} {\bibinfo  {journal} {Nature Physics}\ }\textbf
  {\bibinfo {volume} {16}},\ \bibinfo {pages} {469} (\bibinfo {year}
  {2020})}\BibitemShut {NoStop}%
\bibitem [{\citenamefont {Li}\ \emph {et~al.}(2015)\citenamefont {Li},
  \citenamefont {Liao}, \citenamefont {Zhang}, \citenamefont {Li},
  \citenamefont {Jin}, \citenamefont {Ling}, \citenamefont {Zhang},
  \citenamefont {Zou}, \citenamefont {Pi}, \citenamefont {Yang}, \citenamefont
  {Wang}, \citenamefont {Wu},\ and\ \citenamefont {Zhang}}]{Li2015}%
  \BibitemOpen
  \bibfield  {author} {\bibinfo {author} {\bibfnamefont {Y.}~\bibnamefont
  {Li}}, \bibinfo {author} {\bibfnamefont {H.}~\bibnamefont {Liao}}, \bibinfo
  {author} {\bibfnamefont {Z.}~\bibnamefont {Zhang}}, \bibinfo {author}
  {\bibfnamefont {S.}~\bibnamefont {Li}}, \bibinfo {author} {\bibfnamefont
  {F.}~\bibnamefont {Jin}}, \bibinfo {author} {\bibfnamefont {L.}~\bibnamefont
  {Ling}}, \bibinfo {author} {\bibfnamefont {L.}~\bibnamefont {Zhang}},
  \bibinfo {author} {\bibfnamefont {Y.}~\bibnamefont {Zou}}, \bibinfo {author}
  {\bibfnamefont {L.}~\bibnamefont {Pi}}, \bibinfo {author} {\bibfnamefont
  {Z.}~\bibnamefont {Yang}}, \bibinfo {author} {\bibfnamefont {J.}~\bibnamefont
  {Wang}}, \bibinfo {author} {\bibfnamefont {Z.}~\bibnamefont {Wu}},\ and\
  \bibinfo {author} {\bibfnamefont {Q.}~\bibnamefont {Zhang}},\ }\bibfield
  {title} {\bibinfo {title} {{Gapless quantum spin liquid ground state in the
  two-dimensional spin-1/2 triangular antiferromagnet YbMgGaO$_4$}},\ }\href
  {https://doi.org/10.1038/srep16419} {\bibfield  {journal} {\bibinfo
  {journal} {Scientific Reports}\ }\textbf {\bibinfo {volume} {5}},\ \bibinfo
  {pages} {16419} (\bibinfo {year} {2015})},\ \bibinfo {note}
  {article}\BibitemShut {NoStop}%
\bibitem [{\citenamefont {Li}\ \emph {et~al.}(2017)\citenamefont {Li},
  \citenamefont {Adroja}, \citenamefont {Bewley}, \citenamefont {Voneshen},
  \citenamefont {Tsirlin}, \citenamefont {Gegenwart},\ and\ \citenamefont
  {Zhang}}]{PhysRevLett.118.107202}%
  \BibitemOpen
  \bibfield  {author} {\bibinfo {author} {\bibfnamefont {Y.}~\bibnamefont
  {Li}}, \bibinfo {author} {\bibfnamefont {D.}~\bibnamefont {Adroja}}, \bibinfo
  {author} {\bibfnamefont {R.~I.}\ \bibnamefont {Bewley}}, \bibinfo {author}
  {\bibfnamefont {D.}~\bibnamefont {Voneshen}}, \bibinfo {author}
  {\bibfnamefont {A.~A.}\ \bibnamefont {Tsirlin}}, \bibinfo {author}
  {\bibfnamefont {P.}~\bibnamefont {Gegenwart}},\ and\ \bibinfo {author}
  {\bibfnamefont {Q.}~\bibnamefont {Zhang}},\ }\bibfield  {title} {\bibinfo
  {title} {{Crystalline Electric-Field Randomness in the Triangular Lattice
  Spin-Liquid ${\mathrm{YbMgGaO}}_{4}$}},\ }\href
  {https://doi.org/10.1103/PhysRevLett.118.107202} {\bibfield  {journal}
  {\bibinfo  {journal} {Phys. Rev. Lett.}\ }\textbf {\bibinfo {volume} {118}},\
  \bibinfo {pages} {107202} (\bibinfo {year} {2017})}\BibitemShut {NoStop}%
\bibitem [{\citenamefont {Li}\ \emph {et~al.}(2019)\citenamefont {Li},
  \citenamefont {Bachus}, \citenamefont {Liu}, \citenamefont {Radelytskyi},
  \citenamefont {Bertin}, \citenamefont {Schneidewind}, \citenamefont {Tokiwa},
  \citenamefont {Tsirlin},\ and\ \citenamefont
  {Gegenwart}}]{PhysRevLett.122.137201}%
  \BibitemOpen
  \bibfield  {author} {\bibinfo {author} {\bibfnamefont {Y.}~\bibnamefont
  {Li}}, \bibinfo {author} {\bibfnamefont {S.}~\bibnamefont {Bachus}}, \bibinfo
  {author} {\bibfnamefont {B.}~\bibnamefont {Liu}}, \bibinfo {author}
  {\bibfnamefont {I.}~\bibnamefont {Radelytskyi}}, \bibinfo {author}
  {\bibfnamefont {A.}~\bibnamefont {Bertin}}, \bibinfo {author} {\bibfnamefont
  {A.}~\bibnamefont {Schneidewind}}, \bibinfo {author} {\bibfnamefont
  {Y.}~\bibnamefont {Tokiwa}}, \bibinfo {author} {\bibfnamefont {A.~A.}\
  \bibnamefont {Tsirlin}},\ and\ \bibinfo {author} {\bibfnamefont
  {P.}~\bibnamefont {Gegenwart}},\ }\bibfield  {title} {\bibinfo {title}
  {{Rearrangement of Uncorrelated Valence Bonds Evidenced by Low-Energy Spin
  Excitations in ${\mathrm{YbMgGaO}}_{4}$}},\ }\href
  {https://doi.org/10.1103/PhysRevLett.122.137201} {\bibfield  {journal}
  {\bibinfo  {journal} {Phys. Rev. Lett.}\ }\textbf {\bibinfo {volume} {122}},\
  \bibinfo {pages} {137201} (\bibinfo {year} {2019})}\BibitemShut {NoStop}%
\bibitem [{\citenamefont {Kimchi}\ \emph {et~al.}(2018)\citenamefont {Kimchi},
  \citenamefont {Nahum},\ and\ \citenamefont {Senthil}}]{PhysRevX.8.031028}%
  \BibitemOpen
  \bibfield  {author} {\bibinfo {author} {\bibfnamefont {I.}~\bibnamefont
  {Kimchi}}, \bibinfo {author} {\bibfnamefont {A.}~\bibnamefont {Nahum}},\ and\
  \bibinfo {author} {\bibfnamefont {T.}~\bibnamefont {Senthil}},\ }\bibfield
  {title} {\bibinfo {title} {{Valence Bonds in Random Quantum Magnets: Theory
  and Application to ${\mathrm{YbMgGaO}}_{4}$}},\ }\href
  {https://doi.org/10.1103/PhysRevX.8.031028} {\bibfield  {journal} {\bibinfo
  {journal} {Phys. Rev. X}\ }\textbf {\bibinfo {volume} {8}},\ \bibinfo {pages}
  {031028} (\bibinfo {year} {2018})}\BibitemShut {NoStop}%
\bibitem [{\citenamefont {Takagi}\ \emph {et~al.}(2019)\citenamefont {Takagi},
  \citenamefont {Takayama}, \citenamefont {Jackeli}, \citenamefont
  {Khaliullin},\ and\ \citenamefont {Nagler}}]{Takagi2019}%
  \BibitemOpen
  \bibfield  {author} {\bibinfo {author} {\bibfnamefont {H.}~\bibnamefont
  {Takagi}}, \bibinfo {author} {\bibfnamefont {T.}~\bibnamefont {Takayama}},
  \bibinfo {author} {\bibfnamefont {G.}~\bibnamefont {Jackeli}}, \bibinfo
  {author} {\bibfnamefont {G.}~\bibnamefont {Khaliullin}},\ and\ \bibinfo
  {author} {\bibfnamefont {S.~E.}\ \bibnamefont {Nagler}},\ }\bibfield  {title}
  {\bibinfo {title} {Concept and realization of kitaev quantum spin liquids},\
  }\href {https://doi.org/10.1038/s42254-019-0038-2} {\bibfield  {journal}
  {\bibinfo  {journal} {Nature Reviews Physics}\ }\textbf {\bibinfo {volume}
  {1}},\ \bibinfo {pages} {264} (\bibinfo {year} {2019})}\BibitemShut {NoStop}%
\bibitem [{\citenamefont {Clark}\ \emph {et~al.}(2019)\citenamefont {Clark},
  \citenamefont {Sala}, \citenamefont {Maharaj}, \citenamefont {Stone},
  \citenamefont {Knight}, \citenamefont {Telling}, \citenamefont {Wang},
  \citenamefont {Xu}, \citenamefont {Kim}, \citenamefont {Li}, \citenamefont
  {Cheong},\ and\ \citenamefont {Gaulin}}]{Clark2019}%
  \BibitemOpen
  \bibfield  {author} {\bibinfo {author} {\bibfnamefont {L.}~\bibnamefont
  {Clark}}, \bibinfo {author} {\bibfnamefont {G.}~\bibnamefont {Sala}},
  \bibinfo {author} {\bibfnamefont {D.~D.}\ \bibnamefont {Maharaj}}, \bibinfo
  {author} {\bibfnamefont {M.~B.}\ \bibnamefont {Stone}}, \bibinfo {author}
  {\bibfnamefont {K.~S.}\ \bibnamefont {Knight}}, \bibinfo {author}
  {\bibfnamefont {M.~T.~F.}\ \bibnamefont {Telling}}, \bibinfo {author}
  {\bibfnamefont {X.}~\bibnamefont {Wang}}, \bibinfo {author} {\bibfnamefont
  {X.}~\bibnamefont {Xu}}, \bibinfo {author} {\bibfnamefont {J.}~\bibnamefont
  {Kim}}, \bibinfo {author} {\bibfnamefont {Y.}~\bibnamefont {Li}}, \bibinfo
  {author} {\bibfnamefont {S.-W.}\ \bibnamefont {Cheong}},\ and\ \bibinfo
  {author} {\bibfnamefont {B.~D.}\ \bibnamefont {Gaulin}},\ }\bibfield  {title}
  {\bibinfo {title} {{Two-dimensional spin liquid behaviour in the
  triangular-honeycomb antiferromagnet TbInO$_3$}},\ }\href
  {https://doi.org/10.1038/s41567-018-0407-2} {\bibfield  {journal} {\bibinfo
  {journal} {Nature Physics}\ }\textbf {\bibinfo {volume} {15}},\ \bibinfo
  {pages} {262} (\bibinfo {year} {2019})}\BibitemShut {NoStop}%
\bibitem [{\citenamefont {Arh}\ \emph {et~al.}(2022)\citenamefont {Arh},
  \citenamefont {Sana}, \citenamefont {Pregelj}, \citenamefont {Khuntia},
  \citenamefont {Jagli{\v{c}}i{\'{c}}}, \citenamefont {Le}, \citenamefont
  {Biswas}, \citenamefont {Manuel}, \citenamefont {Mangin-Thro}, \citenamefont
  {Ozarowski},\ and\ \citenamefont {Zorko}}]{Arh2022}%
  \BibitemOpen
  \bibfield  {author} {\bibinfo {author} {\bibfnamefont {T.}~\bibnamefont
  {Arh}}, \bibinfo {author} {\bibfnamefont {B.}~\bibnamefont {Sana}}, \bibinfo
  {author} {\bibfnamefont {M.}~\bibnamefont {Pregelj}}, \bibinfo {author}
  {\bibfnamefont {P.}~\bibnamefont {Khuntia}}, \bibinfo {author} {\bibfnamefont
  {Z.}~\bibnamefont {Jagli{\v{c}}i{\'{c}}}}, \bibinfo {author} {\bibfnamefont
  {M.~D.}\ \bibnamefont {Le}}, \bibinfo {author} {\bibfnamefont {P.~K.}\
  \bibnamefont {Biswas}}, \bibinfo {author} {\bibfnamefont {P.}~\bibnamefont
  {Manuel}}, \bibinfo {author} {\bibfnamefont {L.}~\bibnamefont {Mangin-Thro}},
  \bibinfo {author} {\bibfnamefont {A.}~\bibnamefont {Ozarowski}},\ and\
  \bibinfo {author} {\bibfnamefont {A.}~\bibnamefont {Zorko}},\ }\bibfield
  {title} {\bibinfo {title} {The ising triangular-lattice antiferromagnet
  neodymium heptatantalate as a quantum spin liquid candidate},\ }\href
  {https://doi.org/10.1038/s41563-021-01169-y} {\bibfield  {journal} {\bibinfo
  {journal} {Nature Materials}\ }\textbf {\bibinfo {volume} {21}},\ \bibinfo
  {pages} {416} (\bibinfo {year} {2022})}\BibitemShut {NoStop}%
\bibitem [{\citenamefont {Koteswararao}\ \emph {et~al.}(2014)\citenamefont
  {Koteswararao}, \citenamefont {Kumar}, \citenamefont {Khuntia}, \citenamefont
  {Bhowal}, \citenamefont {Panda}, \citenamefont {Rahman}, \citenamefont
  {Mahajan}, \citenamefont {Dasgupta}, \citenamefont {Baenitz}, \citenamefont
  {Kim},\ and\ \citenamefont {Chou}}]{PhysRevB.90.035141}%
  \BibitemOpen
  \bibfield  {author} {\bibinfo {author} {\bibfnamefont {B.}~\bibnamefont
  {Koteswararao}}, \bibinfo {author} {\bibfnamefont {R.}~\bibnamefont {Kumar}},
  \bibinfo {author} {\bibfnamefont {P.}~\bibnamefont {Khuntia}}, \bibinfo
  {author} {\bibfnamefont {S.}~\bibnamefont {Bhowal}}, \bibinfo {author}
  {\bibfnamefont {S.~K.}\ \bibnamefont {Panda}}, \bibinfo {author}
  {\bibfnamefont {M.~R.}\ \bibnamefont {Rahman}}, \bibinfo {author}
  {\bibfnamefont {A.~V.}\ \bibnamefont {Mahajan}}, \bibinfo {author}
  {\bibfnamefont {I.}~\bibnamefont {Dasgupta}}, \bibinfo {author}
  {\bibfnamefont {M.}~\bibnamefont {Baenitz}}, \bibinfo {author} {\bibfnamefont
  {K.~H.}\ \bibnamefont {Kim}},\ and\ \bibinfo {author} {\bibfnamefont {F.~C.}\
  \bibnamefont {Chou}},\ }\bibfield  {title} {\bibinfo {title} {{Magnetic
  properties and heat capacity of the three-dimensional frustrated
  $S=\frac{1}{2}$ antiferromagnet ${\mathrm{PbCuTe}}_{2}{\mathrm{O}}_{6}$}},\
  }\href {https://doi.org/10.1103/PhysRevB.90.035141} {\bibfield  {journal}
  {\bibinfo  {journal} {Phys. Rev. B}\ }\textbf {\bibinfo {volume} {90}},\
  \bibinfo {pages} {035141} (\bibinfo {year} {2014})}\BibitemShut {NoStop}%
\bibitem [{\citenamefont {Khuntia}\ \emph
  {et~al.}(2016{\natexlab{b}})\citenamefont {Khuntia}, \citenamefont {Bert},
  \citenamefont {Mendels}, \citenamefont {Koteswararao}, \citenamefont
  {Mahajan}, \citenamefont {Baenitz}, \citenamefont {Chou}, \citenamefont
  {Baines}, \citenamefont {Amato},\ and\ \citenamefont
  {Furukawa}}]{PhysRevLett.116.107203}%
  \BibitemOpen
  \bibfield  {author} {\bibinfo {author} {\bibfnamefont {P.}~\bibnamefont
  {Khuntia}}, \bibinfo {author} {\bibfnamefont {F.}~\bibnamefont {Bert}},
  \bibinfo {author} {\bibfnamefont {P.}~\bibnamefont {Mendels}}, \bibinfo
  {author} {\bibfnamefont {B.}~\bibnamefont {Koteswararao}}, \bibinfo {author}
  {\bibfnamefont {A.~V.}\ \bibnamefont {Mahajan}}, \bibinfo {author}
  {\bibfnamefont {M.}~\bibnamefont {Baenitz}}, \bibinfo {author} {\bibfnamefont
  {F.~C.}\ \bibnamefont {Chou}}, \bibinfo {author} {\bibfnamefont
  {C.}~\bibnamefont {Baines}}, \bibinfo {author} {\bibfnamefont
  {A.}~\bibnamefont {Amato}},\ and\ \bibinfo {author} {\bibfnamefont
  {Y.}~\bibnamefont {Furukawa}},\ }\bibfield  {title} {\bibinfo {title} {{Spin
  Liquid State in the 3D Frustrated Antiferromagnet
  ${\mathrm{PbCuTe}}_{2}{\mathrm{O}}_{6}$: NMR and Muon Spin Relaxation
  Studies}},\ }\href {https://doi.org/10.1103/PhysRevLett.116.107203}
  {\bibfield  {journal} {\bibinfo  {journal} {Phys. Rev. Lett.}\ }\textbf
  {\bibinfo {volume} {116}},\ \bibinfo {pages} {107203} (\bibinfo {year}
  {2016}{\natexlab{b}})}\BibitemShut {NoStop}%
\bibitem [{\citenamefont {Chillal}\ \emph {et~al.}(2020)\citenamefont
  {Chillal}, \citenamefont {Iqbal}, \citenamefont {Jeschke}, \citenamefont
  {Rodriguez-Rivera}, \citenamefont {Bewley}, \citenamefont {Manuel},
  \citenamefont {Khalyavin}, \citenamefont {Steffens}, \citenamefont {Thomale},
  \citenamefont {Islam}, \citenamefont {Reuther},\ and\ \citenamefont
  {Lake}}]{Chillal2020}%
  \BibitemOpen
  \bibfield  {author} {\bibinfo {author} {\bibfnamefont {S.}~\bibnamefont
  {Chillal}}, \bibinfo {author} {\bibfnamefont {Y.}~\bibnamefont {Iqbal}},
  \bibinfo {author} {\bibfnamefont {H.~O.}\ \bibnamefont {Jeschke}}, \bibinfo
  {author} {\bibfnamefont {J.~A.}\ \bibnamefont {Rodriguez-Rivera}}, \bibinfo
  {author} {\bibfnamefont {R.}~\bibnamefont {Bewley}}, \bibinfo {author}
  {\bibfnamefont {P.}~\bibnamefont {Manuel}}, \bibinfo {author} {\bibfnamefont
  {D.}~\bibnamefont {Khalyavin}}, \bibinfo {author} {\bibfnamefont
  {P.}~\bibnamefont {Steffens}}, \bibinfo {author} {\bibfnamefont
  {R.}~\bibnamefont {Thomale}}, \bibinfo {author} {\bibfnamefont {A.~T. M.~N.}\
  \bibnamefont {Islam}}, \bibinfo {author} {\bibfnamefont {J.}~\bibnamefont
  {Reuther}},\ and\ \bibinfo {author} {\bibfnamefont {B.}~\bibnamefont
  {Lake}},\ }\bibfield  {title} {\bibinfo {title} {{Evidence for a
  three-dimensional quantum spin liquid in PbCuTe$_2$O$_6$}},\ }\href
  {https://doi.org/10.1038/s41467-020-15594-1} {\bibfield  {journal} {\bibinfo
  {journal} {Nature Communications}\ }\textbf {\bibinfo {volume} {11}},\
  \bibinfo {pages} {2348} (\bibinfo {year} {2020})}\BibitemShut {NoStop}%
\bibitem [{\citenamefont {Balz}\ \emph {et~al.}(2016)\citenamefont {Balz},
  \citenamefont {Lake}, \citenamefont {Reuther}, \citenamefont {Luetkens},
  \citenamefont {Sch{\"o}nemann}, \citenamefont {Herrmannsd{\"o}rfer},
  \citenamefont {Singh}, \citenamefont {Nazmul~Islam}, \citenamefont {Wheeler},
  \citenamefont {Rodriguez-Rivera}, \citenamefont {Guidi}, \citenamefont
  {Simeoni}, \citenamefont {Baines},\ and\ \citenamefont {Ryll}}]{Balz2016}%
  \BibitemOpen
  \bibfield  {author} {\bibinfo {author} {\bibfnamefont {C.}~\bibnamefont
  {Balz}}, \bibinfo {author} {\bibfnamefont {B.}~\bibnamefont {Lake}}, \bibinfo
  {author} {\bibfnamefont {J.}~\bibnamefont {Reuther}}, \bibinfo {author}
  {\bibfnamefont {H.}~\bibnamefont {Luetkens}}, \bibinfo {author}
  {\bibfnamefont {R.}~\bibnamefont {Sch{\"o}nemann}}, \bibinfo {author}
  {\bibfnamefont {T.}~\bibnamefont {Herrmannsd{\"o}rfer}}, \bibinfo {author}
  {\bibfnamefont {Y.}~\bibnamefont {Singh}}, \bibinfo {author} {\bibfnamefont
  {A.~T.~M.}\ \bibnamefont {Nazmul~Islam}}, \bibinfo {author} {\bibfnamefont
  {E.~M.}\ \bibnamefont {Wheeler}}, \bibinfo {author} {\bibfnamefont
  {J.}~\bibnamefont {Rodriguez-Rivera}}, \bibinfo {author} {\bibfnamefont
  {T.}~\bibnamefont {Guidi}}, \bibinfo {author} {\bibfnamefont
  {G.}~\bibnamefont {Simeoni}}, \bibinfo {author} {\bibfnamefont
  {C.}~\bibnamefont {Baines}},\ and\ \bibinfo {author} {\bibfnamefont
  {H.}~\bibnamefont {Ryll}},\ }\bibfield  {title} {\bibinfo {title} {Physical
  realization of a quantum spin liquid based on a complex frustration
  mechanism},\ }\href {https://doi.org/10.1038/nphys3826} {\bibfield  {journal}
  {\bibinfo  {journal} {Nature Physics}\ }\textbf {\bibinfo {volume} {12}},\
  \bibinfo {pages} {942} (\bibinfo {year} {2016})}\BibitemShut {NoStop}%
\bibitem [{\citenamefont {Balz}\ \emph {et~al.}(2017)\citenamefont {Balz},
  \citenamefont {Lake}, \citenamefont {Nazmul~Islam}, \citenamefont {Singh},
  \citenamefont {Rodriguez-Rivera}, \citenamefont {Guidi}, \citenamefont
  {Wheeler}, \citenamefont {Simeoni},\ and\ \citenamefont
  {Ryll}}]{PhysRevB.95.174414}%
  \BibitemOpen
  \bibfield  {author} {\bibinfo {author} {\bibfnamefont {C.}~\bibnamefont
  {Balz}}, \bibinfo {author} {\bibfnamefont {B.}~\bibnamefont {Lake}}, \bibinfo
  {author} {\bibfnamefont {A.~T.~M.}\ \bibnamefont {Nazmul~Islam}}, \bibinfo
  {author} {\bibfnamefont {Y.}~\bibnamefont {Singh}}, \bibinfo {author}
  {\bibfnamefont {J.~A.}\ \bibnamefont {Rodriguez-Rivera}}, \bibinfo {author}
  {\bibfnamefont {T.}~\bibnamefont {Guidi}}, \bibinfo {author} {\bibfnamefont
  {E.~M.}\ \bibnamefont {Wheeler}}, \bibinfo {author} {\bibfnamefont {G.~G.}\
  \bibnamefont {Simeoni}},\ and\ \bibinfo {author} {\bibfnamefont
  {H.}~\bibnamefont {Ryll}},\ }\bibfield  {title} {\bibinfo {title} {{Magnetic
  Hamiltonian and phase diagram of the quantum spin liquid
  ${\mathrm{Ca}}_{10}{\mathrm{Cr}}_{7}{\mathrm{O}}_{28}$}},\ }\href
  {https://doi.org/10.1103/PhysRevB.95.174414} {\bibfield  {journal} {\bibinfo
  {journal} {Phys. Rev. B}\ }\textbf {\bibinfo {volume} {95}},\ \bibinfo
  {pages} {174414} (\bibinfo {year} {2017})}\BibitemShut {NoStop}%
\bibitem [{\citenamefont {Plumb}\ \emph {et~al.}(2019)\citenamefont {Plumb},
  \citenamefont {Changlani}, \citenamefont {Scheie}, \citenamefont {Zhang},
  \citenamefont {Krizan}, \citenamefont {Rodriguez-Rivera}, \citenamefont
  {Qiu}, \citenamefont {Winn}, \citenamefont {Cava},\ and\ \citenamefont
  {Broholm}}]{Plumb2019}%
  \BibitemOpen
  \bibfield  {author} {\bibinfo {author} {\bibfnamefont {K.~W.}\ \bibnamefont
  {Plumb}}, \bibinfo {author} {\bibfnamefont {H.~J.}\ \bibnamefont
  {Changlani}}, \bibinfo {author} {\bibfnamefont {A.}~\bibnamefont {Scheie}},
  \bibinfo {author} {\bibfnamefont {S.}~\bibnamefont {Zhang}}, \bibinfo
  {author} {\bibfnamefont {J.~W.}\ \bibnamefont {Krizan}}, \bibinfo {author}
  {\bibfnamefont {J.~A.}\ \bibnamefont {Rodriguez-Rivera}}, \bibinfo {author}
  {\bibfnamefont {Y.}~\bibnamefont {Qiu}}, \bibinfo {author} {\bibfnamefont
  {B.}~\bibnamefont {Winn}}, \bibinfo {author} {\bibfnamefont {R.~J.}\
  \bibnamefont {Cava}},\ and\ \bibinfo {author} {\bibfnamefont {C.~L.}\
  \bibnamefont {Broholm}},\ }\bibfield  {title} {\bibinfo {title} {{Continuum
  of quantum fluctuations in a three-dimensional S{\thinspace}={\thinspace}1
  Heisenberg magnet}},\ }\href {https://doi.org/10.1038/s41567-018-0317-3}
  {\bibfield  {journal} {\bibinfo  {journal} {Nature Physics}\ }\textbf
  {\bibinfo {volume} {15}},\ \bibinfo {pages} {54} (\bibinfo {year}
  {2019})}\BibitemShut {NoStop}%
\bibitem [{\citenamefont {Gao}\ \emph {et~al.}(2019)\citenamefont {Gao},
  \citenamefont {Chen}, \citenamefont {Tam}, \citenamefont {Huang},
  \citenamefont {Sasmal}, \citenamefont {Adroja}, \citenamefont {Ye},
  \citenamefont {Cao}, \citenamefont {Sala}, \citenamefont {Stone},
  \citenamefont {Baines}, \citenamefont {Verezhak}, \citenamefont {Hu},
  \citenamefont {Chung}, \citenamefont {Xu}, \citenamefont {Cheong},
  \citenamefont {Nallaiyan}, \citenamefont {Spagna}, \citenamefont {Maple},
  \citenamefont {Nevidomskyy}, \citenamefont {Morosan}, \citenamefont {Chen},\
  and\ \citenamefont {Dai}}]{Gao2019}%
  \BibitemOpen
  \bibfield  {author} {\bibinfo {author} {\bibfnamefont {B.}~\bibnamefont
  {Gao}}, \bibinfo {author} {\bibfnamefont {T.}~\bibnamefont {Chen}}, \bibinfo
  {author} {\bibfnamefont {D.~W.}\ \bibnamefont {Tam}}, \bibinfo {author}
  {\bibfnamefont {C.-L.}\ \bibnamefont {Huang}}, \bibinfo {author}
  {\bibfnamefont {K.}~\bibnamefont {Sasmal}}, \bibinfo {author} {\bibfnamefont
  {D.~T.}\ \bibnamefont {Adroja}}, \bibinfo {author} {\bibfnamefont
  {F.}~\bibnamefont {Ye}}, \bibinfo {author} {\bibfnamefont {H.}~\bibnamefont
  {Cao}}, \bibinfo {author} {\bibfnamefont {G.}~\bibnamefont {Sala}}, \bibinfo
  {author} {\bibfnamefont {M.~B.}\ \bibnamefont {Stone}}, \bibinfo {author}
  {\bibfnamefont {C.}~\bibnamefont {Baines}}, \bibinfo {author} {\bibfnamefont
  {J.~A.~T.}\ \bibnamefont {Verezhak}}, \bibinfo {author} {\bibfnamefont
  {H.}~\bibnamefont {Hu}}, \bibinfo {author} {\bibfnamefont {J.-H.}\
  \bibnamefont {Chung}}, \bibinfo {author} {\bibfnamefont {X.}~\bibnamefont
  {Xu}}, \bibinfo {author} {\bibfnamefont {S.-W.}\ \bibnamefont {Cheong}},
  \bibinfo {author} {\bibfnamefont {M.}~\bibnamefont {Nallaiyan}}, \bibinfo
  {author} {\bibfnamefont {S.}~\bibnamefont {Spagna}}, \bibinfo {author}
  {\bibfnamefont {M.~B.}\ \bibnamefont {Maple}}, \bibinfo {author}
  {\bibfnamefont {A.~H.}\ \bibnamefont {Nevidomskyy}}, \bibinfo {author}
  {\bibfnamefont {E.}~\bibnamefont {Morosan}}, \bibinfo {author} {\bibfnamefont
  {G.}~\bibnamefont {Chen}},\ and\ \bibinfo {author} {\bibfnamefont
  {P.}~\bibnamefont {Dai}},\ }\bibfield  {title} {\bibinfo {title}
  {{Experimental signatures of a three-dimensional quantum spin liquid in
  effective spin-1/2 Ce$_2$Zr$_2$O$_7$ pyrochlore}},\ }\href
  {https://doi.org/10.1038/s41567-019-0577-6} {\bibfield  {journal} {\bibinfo
  {journal} {Nature Physics}\ }\textbf {\bibinfo {volume} {15}},\ \bibinfo
  {pages} {1052} (\bibinfo {year} {2019})}\BibitemShut {NoStop}%
\bibitem [{\citenamefont {Gaudet}\ \emph {et~al.}(2019)\citenamefont {Gaudet},
  \citenamefont {Smith}, \citenamefont {Dudemaine}, \citenamefont {Beare},
  \citenamefont {Buhariwalla}, \citenamefont {Butch}, \citenamefont {Stone},
  \citenamefont {Kolesnikov}, \citenamefont {Xu}, \citenamefont {Yahne},
  \citenamefont {Ross}, \citenamefont {Marjerrison}, \citenamefont {Garrett},
  \citenamefont {Luke}, \citenamefont {Bianchi},\ and\ \citenamefont
  {Gaulin}}]{PhysRevLett.122.187201}%
  \BibitemOpen
  \bibfield  {author} {\bibinfo {author} {\bibfnamefont {J.}~\bibnamefont
  {Gaudet}}, \bibinfo {author} {\bibfnamefont {E.~M.}\ \bibnamefont {Smith}},
  \bibinfo {author} {\bibfnamefont {J.}~\bibnamefont {Dudemaine}}, \bibinfo
  {author} {\bibfnamefont {J.}~\bibnamefont {Beare}}, \bibinfo {author}
  {\bibfnamefont {C.~R.~C.}\ \bibnamefont {Buhariwalla}}, \bibinfo {author}
  {\bibfnamefont {N.~P.}\ \bibnamefont {Butch}}, \bibinfo {author}
  {\bibfnamefont {M.~B.}\ \bibnamefont {Stone}}, \bibinfo {author}
  {\bibfnamefont {A.~I.}\ \bibnamefont {Kolesnikov}}, \bibinfo {author}
  {\bibfnamefont {G.}~\bibnamefont {Xu}}, \bibinfo {author} {\bibfnamefont
  {D.~R.}\ \bibnamefont {Yahne}}, \bibinfo {author} {\bibfnamefont {K.~A.}\
  \bibnamefont {Ross}}, \bibinfo {author} {\bibfnamefont {C.~A.}\ \bibnamefont
  {Marjerrison}}, \bibinfo {author} {\bibfnamefont {J.~D.}\ \bibnamefont
  {Garrett}}, \bibinfo {author} {\bibfnamefont {G.~M.}\ \bibnamefont {Luke}},
  \bibinfo {author} {\bibfnamefont {A.~D.}\ \bibnamefont {Bianchi}},\ and\
  \bibinfo {author} {\bibfnamefont {B.~D.}\ \bibnamefont {Gaulin}},\ }\bibfield
   {title} {\bibinfo {title} {{Quantum Spin Ice Dynamics in the Dipole-Octupole
  Pyrochlore Magnet ${\mathrm{Ce}}_{2}{\mathrm{Zr}}_{2}{\mathrm{O}}_{7}$}},\
  }\href {https://doi.org/10.1103/PhysRevLett.122.187201} {\bibfield  {journal}
  {\bibinfo  {journal} {Phys. Rev. Lett.}\ }\textbf {\bibinfo {volume} {122}},\
  \bibinfo {pages} {187201} (\bibinfo {year} {2019})}\BibitemShut {NoStop}%
\bibitem [{\citenamefont {Li}\ and\ \citenamefont
  {Chen}(2017)}]{PhysRevB.95.041106}%
  \BibitemOpen
  \bibfield  {author} {\bibinfo {author} {\bibfnamefont {Y.-D.}\ \bibnamefont
  {Li}}\ and\ \bibinfo {author} {\bibfnamefont {G.}~\bibnamefont {Chen}},\
  }\bibfield  {title} {\bibinfo {title} {Symmetry enriched u(1) topological
  orders for dipole-octupole doublets on a pyrochlore lattice},\ }\href
  {https://doi.org/10.1103/PhysRevB.95.041106} {\bibfield  {journal} {\bibinfo
  {journal} {Phys. Rev. B}\ }\textbf {\bibinfo {volume} {95}},\ \bibinfo
  {pages} {041106} (\bibinfo {year} {2017})}\BibitemShut {NoStop}%
\bibitem [{\citenamefont {Smith}\ \emph {et~al.}(2022)\citenamefont {Smith},
  \citenamefont {Benton}, \citenamefont {Yahne}, \citenamefont {Placke},
  \citenamefont {Sch\"afer}, \citenamefont {Gaudet}, \citenamefont {Dudemaine},
  \citenamefont {Fitterman}, \citenamefont {Beare}, \citenamefont {Wildes},
  \citenamefont {Bhattacharya}, \citenamefont {DeLazzer}, \citenamefont
  {Buhariwalla}, \citenamefont {Butch}, \citenamefont {Movshovich},
  \citenamefont {Garrett}, \citenamefont {Marjerrison}, \citenamefont {Clancy},
  \citenamefont {Kermarrec}, \citenamefont {Luke}, \citenamefont {Bianchi},
  \citenamefont {Ross},\ and\ \citenamefont {Gaulin}}]{PhysRevX.12.021015}%
  \BibitemOpen
  \bibfield  {author} {\bibinfo {author} {\bibfnamefont {E.~M.}\ \bibnamefont
  {Smith}}, \bibinfo {author} {\bibfnamefont {O.}~\bibnamefont {Benton}},
  \bibinfo {author} {\bibfnamefont {D.~R.}\ \bibnamefont {Yahne}}, \bibinfo
  {author} {\bibfnamefont {B.}~\bibnamefont {Placke}}, \bibinfo {author}
  {\bibfnamefont {R.}~\bibnamefont {Sch\"afer}}, \bibinfo {author}
  {\bibfnamefont {J.}~\bibnamefont {Gaudet}}, \bibinfo {author} {\bibfnamefont
  {J.}~\bibnamefont {Dudemaine}}, \bibinfo {author} {\bibfnamefont
  {A.}~\bibnamefont {Fitterman}}, \bibinfo {author} {\bibfnamefont
  {J.}~\bibnamefont {Beare}}, \bibinfo {author} {\bibfnamefont {A.~R.}\
  \bibnamefont {Wildes}}, \bibinfo {author} {\bibfnamefont {S.}~\bibnamefont
  {Bhattacharya}}, \bibinfo {author} {\bibfnamefont {T.}~\bibnamefont
  {DeLazzer}}, \bibinfo {author} {\bibfnamefont {C.~R.~C.}\ \bibnamefont
  {Buhariwalla}}, \bibinfo {author} {\bibfnamefont {N.~P.}\ \bibnamefont
  {Butch}}, \bibinfo {author} {\bibfnamefont {R.}~\bibnamefont {Movshovich}},
  \bibinfo {author} {\bibfnamefont {J.~D.}\ \bibnamefont {Garrett}}, \bibinfo
  {author} {\bibfnamefont {C.~A.}\ \bibnamefont {Marjerrison}}, \bibinfo
  {author} {\bibfnamefont {J.~P.}\ \bibnamefont {Clancy}}, \bibinfo {author}
  {\bibfnamefont {E.}~\bibnamefont {Kermarrec}}, \bibinfo {author}
  {\bibfnamefont {G.~M.}\ \bibnamefont {Luke}}, \bibinfo {author}
  {\bibfnamefont {A.~D.}\ \bibnamefont {Bianchi}}, \bibinfo {author}
  {\bibfnamefont {K.~A.}\ \bibnamefont {Ross}},\ and\ \bibinfo {author}
  {\bibfnamefont {B.~D.}\ \bibnamefont {Gaulin}},\ }\bibfield  {title}
  {\bibinfo {title} {{Case for a ${\mathrm{U}(1)}_{\ensuremath{\pi}}$ Quantum
  Spin Liquid Ground State in the Dipole-Octupole Pyrochlore
  ${\mathrm{Ce}}_{2}{\mathrm{Zr}}_{2}{\mathrm{O}}_{7}$}},\ }\href
  {https://doi.org/10.1103/PhysRevX.12.021015} {\bibfield  {journal} {\bibinfo
  {journal} {Phys. Rev. X}\ }\textbf {\bibinfo {volume} {12}},\ \bibinfo
  {pages} {021015} (\bibinfo {year} {2022})}\BibitemShut {NoStop}%
\bibitem [{\citenamefont {Okamoto}\ \emph {et~al.}(2007)\citenamefont
  {Okamoto}, \citenamefont {Nohara}, \citenamefont {Aruga-Katori},\ and\
  \citenamefont {Takagi}}]{PhysRevLett.99.137207}%
  \BibitemOpen
  \bibfield  {author} {\bibinfo {author} {\bibfnamefont {Y.}~\bibnamefont
  {Okamoto}}, \bibinfo {author} {\bibfnamefont {M.}~\bibnamefont {Nohara}},
  \bibinfo {author} {\bibfnamefont {H.}~\bibnamefont {Aruga-Katori}},\ and\
  \bibinfo {author} {\bibfnamefont {H.}~\bibnamefont {Takagi}},\ }\bibfield
  {title} {\bibinfo {title} {{Spin-Liquid State in the $S=1/2$ Hyperkagome
  Antiferromagnet ${\mathrm{Na}}_{4}{\mathrm{Ir}}_{3}{\mathrm{O}}_{8}$}},\
  }\href {https://doi.org/10.1103/PhysRevLett.99.137207} {\bibfield  {journal}
  {\bibinfo  {journal} {Phys. Rev. Lett.}\ }\textbf {\bibinfo {volume} {99}},\
  \bibinfo {pages} {137207} (\bibinfo {year} {2007})}\BibitemShut {NoStop}%
\bibitem [{\citenamefont {Zhou}\ \emph {et~al.}(2008)\citenamefont {Zhou},
  \citenamefont {Lee}, \citenamefont {Ng},\ and\ \citenamefont
  {Zhang}}]{PhysRevLett.101.197201}%
  \BibitemOpen
  \bibfield  {author} {\bibinfo {author} {\bibfnamefont {Y.}~\bibnamefont
  {Zhou}}, \bibinfo {author} {\bibfnamefont {P.~A.}\ \bibnamefont {Lee}},
  \bibinfo {author} {\bibfnamefont {T.-K.}\ \bibnamefont {Ng}},\ and\ \bibinfo
  {author} {\bibfnamefont {F.-C.}\ \bibnamefont {Zhang}},\ }\bibfield  {title}
  {\bibinfo {title} {{${\mathrm{Na}}_{4}{\mathrm{Ir}}_{3}{\mathrm{O}}_{8}$ as a
  3D Spin Liquid with Fermionic Spinons}},\ }\href
  {https://doi.org/10.1103/PhysRevLett.101.197201} {\bibfield  {journal}
  {\bibinfo  {journal} {Phys. Rev. Lett.}\ }\textbf {\bibinfo {volume} {101}},\
  \bibinfo {pages} {197201} (\bibinfo {year} {2008})}\BibitemShut {NoStop}%
\bibitem [{\citenamefont {Lawler}\ \emph {et~al.}(2008)\citenamefont {Lawler},
  \citenamefont {Kee}, \citenamefont {Kim},\ and\ \citenamefont
  {Vishwanath}}]{PhysRevLett.100.227201}%
  \BibitemOpen
  \bibfield  {author} {\bibinfo {author} {\bibfnamefont {M.~J.}\ \bibnamefont
  {Lawler}}, \bibinfo {author} {\bibfnamefont {H.-Y.}\ \bibnamefont {Kee}},
  \bibinfo {author} {\bibfnamefont {Y.~B.}\ \bibnamefont {Kim}},\ and\ \bibinfo
  {author} {\bibfnamefont {A.}~\bibnamefont {Vishwanath}},\ }\bibfield  {title}
  {\bibinfo {title} {{Topological Spin Liquid on the Hyperkagome Lattice of
  ${\mathrm{Na}}_{4}{\mathrm{Ir}}_{3}{\mathrm{O}}_{8}$}},\ }\href
  {https://doi.org/10.1103/PhysRevLett.100.227201} {\bibfield  {journal}
  {\bibinfo  {journal} {Phys. Rev. Lett.}\ }\textbf {\bibinfo {volume} {100}},\
  \bibinfo {pages} {227201} (\bibinfo {year} {2008})}\BibitemShut {NoStop}%
\bibitem [{\citenamefont {Shockley}\ \emph {et~al.}(2015)\citenamefont
  {Shockley}, \citenamefont {Bert}, \citenamefont {Orain}, \citenamefont
  {Okamoto},\ and\ \citenamefont {Mendels}}]{PhysRevLett.115.047201}%
  \BibitemOpen
  \bibfield  {author} {\bibinfo {author} {\bibfnamefont {A.~C.}\ \bibnamefont
  {Shockley}}, \bibinfo {author} {\bibfnamefont {F.}~\bibnamefont {Bert}},
  \bibinfo {author} {\bibfnamefont {J.-C.}\ \bibnamefont {Orain}}, \bibinfo
  {author} {\bibfnamefont {Y.}~\bibnamefont {Okamoto}},\ and\ \bibinfo {author}
  {\bibfnamefont {P.}~\bibnamefont {Mendels}},\ }\bibfield  {title} {\bibinfo
  {title} {{Frozen State and Spin Liquid Physics in
  ${\mathrm{Na}}_{4}{\mathrm{Ir}}_{3}{\mathrm{O}}_{8}$: An NMR Study}},\ }\href
  {https://doi.org/10.1103/PhysRevLett.115.047201} {\bibfield  {journal}
  {\bibinfo  {journal} {Phys. Rev. Lett.}\ }\textbf {\bibinfo {volume} {115}},\
  \bibinfo {pages} {047201} (\bibinfo {year} {2015})}\BibitemShut {NoStop}%
\bibitem [{\citenamefont {Khomskii}(2014)}]{khomskii_2014}%
  \BibitemOpen
  \bibfield  {author} {\bibinfo {author} {\bibfnamefont {D.~I.}\ \bibnamefont
  {Khomskii}},\ }\href {https://doi.org/10.1017/CBO9781139096782} {\emph
  {\bibinfo {title} {Transition Metal Compounds}}}\ (\bibinfo  {publisher}
  {Cambridge University Press},\ \bibinfo {year} {2014})\BibitemShut {NoStop}%
\bibitem [{\citenamefont {Khatua}\ \emph {et~al.}(2022)\citenamefont {Khatua},
  \citenamefont {Bhattacharya}, \citenamefont {Ding}, \citenamefont {Vrtnik},
  \citenamefont {Strydom}, \citenamefont {Butch}, \citenamefont {Luetkens},
  \citenamefont {Kermarrec}, \citenamefont {Rao}, \citenamefont {Zorko},
  \citenamefont {Furukawa},\ and\ \citenamefont
  {Khuntia}}]{PhysRevB.106.104404}%
  \BibitemOpen
  \bibfield  {author} {\bibinfo {author} {\bibfnamefont {J.}~\bibnamefont
  {Khatua}}, \bibinfo {author} {\bibfnamefont {S.}~\bibnamefont
  {Bhattacharya}}, \bibinfo {author} {\bibfnamefont {Q.~P.}\ \bibnamefont
  {Ding}}, \bibinfo {author} {\bibfnamefont {S.}~\bibnamefont {Vrtnik}},
  \bibinfo {author} {\bibfnamefont {A.~M.}\ \bibnamefont {Strydom}}, \bibinfo
  {author} {\bibfnamefont {N.~P.}\ \bibnamefont {Butch}}, \bibinfo {author}
  {\bibfnamefont {H.}~\bibnamefont {Luetkens}}, \bibinfo {author}
  {\bibfnamefont {E.}~\bibnamefont {Kermarrec}}, \bibinfo {author}
  {\bibfnamefont {M.~S.~R.}\ \bibnamefont {Rao}}, \bibinfo {author}
  {\bibfnamefont {A.}~\bibnamefont {Zorko}}, \bibinfo {author} {\bibfnamefont
  {Y.}~\bibnamefont {Furukawa}},\ and\ \bibinfo {author} {\bibfnamefont
  {P.}~\bibnamefont {Khuntia}},\ }\bibfield  {title} {\bibinfo {title} {{Spin
  liquid state in a rare-earth hyperkagome lattice}},\ }\href
  {https://doi.org/10.1103/PhysRevB.106.104404} {\bibfield  {journal} {\bibinfo
   {journal} {Phys. Rev. B}\ }\textbf {\bibinfo {volume} {106}},\ \bibinfo
  {pages} {104404} (\bibinfo {year} {2022})}\BibitemShut {NoStop}%
\bibitem [{\citenamefont {Mukherjee}\ \emph {et~al.}(2017)\citenamefont
  {Mukherjee}, \citenamefont {Hamilton}, \citenamefont {Glass},\ and\
  \citenamefont {Dutton}}]{Mukherjee_2017}%
  \BibitemOpen
  \bibfield  {author} {\bibinfo {author} {\bibfnamefont {P.}~\bibnamefont
  {Mukherjee}}, \bibinfo {author} {\bibfnamefont {A.~C.~S.}\ \bibnamefont
  {Hamilton}}, \bibinfo {author} {\bibfnamefont {H.~F.~J.}\ \bibnamefont
  {Glass}},\ and\ \bibinfo {author} {\bibfnamefont {S.~E.}\ \bibnamefont
  {Dutton}},\ }\bibfield  {title} {\bibinfo {title} {{Sensitivity of magnetic
  properties to chemical pressure in lanthanide {garnets
  Ln}$_3$A$_2$X$_3$O$_{12}$,Ln{\hspace{0.167em}}{\hspace{0.167em}}={\hspace{0.167em}}{\hspace{0.167em}}Gd,
  Tb, Dy,
  Ho,A{\hspace{0.167em}}{\hspace{0.167em}}={\hspace{0.167em}}{\hspace{0.167em}}Ga,
  Sc, In,
  Te,X{\hspace{0.167em}}{\hspace{0.167em}}={\hspace{0.167em}}{\hspace{0.167em}}Ga,
  Al, Li}},\ }\href {https://doi.org/10.1088/1361-648x/aa810e} {\bibfield
  {journal} {\bibinfo  {journal} {Journal of Physics: Condensed Matter}\
  }\textbf {\bibinfo {volume} {29}},\ \bibinfo {pages} {405808} (\bibinfo
  {year} {2017})}\BibitemShut {NoStop}%
\bibitem [{\citenamefont {Paddison}\ \emph {et~al.}(2015)\citenamefont
  {Paddison}, \citenamefont {Jacobsen}, \citenamefont {Petrenko}, \citenamefont
  {Fernández-Díaz}, \citenamefont {Deen},\ and\ \citenamefont
  {Goodwin}}]{doi:10.1126/science.aaa5326}%
  \BibitemOpen
  \bibfield  {author} {\bibinfo {author} {\bibfnamefont {J.~A.~M.}\
  \bibnamefont {Paddison}}, \bibinfo {author} {\bibfnamefont {H.}~\bibnamefont
  {Jacobsen}}, \bibinfo {author} {\bibfnamefont {O.~A.}\ \bibnamefont
  {Petrenko}}, \bibinfo {author} {\bibfnamefont {M.~T.}\ \bibnamefont
  {Fernández-Díaz}}, \bibinfo {author} {\bibfnamefont {P.~P.}\ \bibnamefont
  {Deen}},\ and\ \bibinfo {author} {\bibfnamefont {A.~L.}\ \bibnamefont
  {Goodwin}},\ }\bibfield  {title} {\bibinfo {title} {{Hidden order in
  spin-liquid Gd$_3$Ga$_5$O$_{12}$}},\ }\href
  {https://doi.org/10.1126/science.aaa5326} {\bibfield  {journal} {\bibinfo
  {journal} {Science}\ }\textbf {\bibinfo {volume} {350}},\ \bibinfo {pages}
  {179} (\bibinfo {year} {2015})}\BibitemShut {NoStop}%
\bibitem [{\citenamefont {Jacobsen}\ \emph {et~al.}(2021)\citenamefont
  {Jacobsen}, \citenamefont {Florea}, \citenamefont {Lhotel}, \citenamefont
  {Lefmann}, \citenamefont {Petrenko}, \citenamefont {Knee}, \citenamefont
  {Seydel}, \citenamefont {Henry}, \citenamefont {Bewley}, \citenamefont
  {Voneshen}, \citenamefont {Wildes}, \citenamefont {Nilsen},\ and\
  \citenamefont {Deen}}]{PhysRevB.104.054440}%
  \BibitemOpen
  \bibfield  {author} {\bibinfo {author} {\bibfnamefont {H.}~\bibnamefont
  {Jacobsen}}, \bibinfo {author} {\bibfnamefont {O.}~\bibnamefont {Florea}},
  \bibinfo {author} {\bibfnamefont {E.}~\bibnamefont {Lhotel}}, \bibinfo
  {author} {\bibfnamefont {K.}~\bibnamefont {Lefmann}}, \bibinfo {author}
  {\bibfnamefont {O.~A.}\ \bibnamefont {Petrenko}}, \bibinfo {author}
  {\bibfnamefont {C.~S.}\ \bibnamefont {Knee}}, \bibinfo {author}
  {\bibfnamefont {T.}~\bibnamefont {Seydel}}, \bibinfo {author} {\bibfnamefont
  {P.~F.}\ \bibnamefont {Henry}}, \bibinfo {author} {\bibfnamefont
  {R.}~\bibnamefont {Bewley}}, \bibinfo {author} {\bibfnamefont
  {D.}~\bibnamefont {Voneshen}}, \bibinfo {author} {\bibfnamefont
  {A.}~\bibnamefont {Wildes}}, \bibinfo {author} {\bibfnamefont
  {G.}~\bibnamefont {Nilsen}},\ and\ \bibinfo {author} {\bibfnamefont {P.~P.}\
  \bibnamefont {Deen}},\ }\bibfield  {title} {\bibinfo {title} {Spin dynamics
  of the director state in frustrated hyperkagome systems},\ }\href
  {https://doi.org/10.1103/PhysRevB.104.054440} {\bibfield  {journal} {\bibinfo
   {journal} {Phys. Rev. B}\ }\textbf {\bibinfo {volume} {104}},\ \bibinfo
  {pages} {054440} (\bibinfo {year} {2021})}\BibitemShut {NoStop}%
\bibitem [{\citenamefont {Machida}\ \emph {et~al.}(2010)\citenamefont
  {Machida}, \citenamefont {Nakatsuji}, \citenamefont {Onoda}, \citenamefont
  {Tayama},\ and\ \citenamefont {Sakakibara}}]{Machida2010}%
  \BibitemOpen
  \bibfield  {author} {\bibinfo {author} {\bibfnamefont {Y.}~\bibnamefont
  {Machida}}, \bibinfo {author} {\bibfnamefont {S.}~\bibnamefont {Nakatsuji}},
  \bibinfo {author} {\bibfnamefont {S.}~\bibnamefont {Onoda}}, \bibinfo
  {author} {\bibfnamefont {T.}~\bibnamefont {Tayama}},\ and\ \bibinfo {author}
  {\bibfnamefont {T.}~\bibnamefont {Sakakibara}},\ }\bibfield  {title}
  {\bibinfo {title} {Time-reversal symmetry breaking and spontaneous hall
  effect without magnetic dipole order},\ }\href
  {https://doi.org/10.1038/nature08680} {\bibfield  {journal} {\bibinfo
  {journal} {Nature}\ }\textbf {\bibinfo {volume} {463}},\ \bibinfo {pages}
  {210} (\bibinfo {year} {2010})}\BibitemShut {NoStop}%
\bibitem [{\citenamefont {Sandberg}\ \emph
  {et~al.}(2021{\natexlab{a}})\citenamefont {Sandberg}, \citenamefont {Edberg},
  \citenamefont {Bakke}, \citenamefont {Pedersen}, \citenamefont {Hatnean},
  \citenamefont {Balakrishnan}, \citenamefont {Mangin-Thro}, \citenamefont
  {Wildes}, \citenamefont {F\aa{}k}, \citenamefont {Ehlers}, \citenamefont
  {Sala}, \citenamefont {Henelius}, \citenamefont {Lefmann},\ and\
  \citenamefont {Deen}}]{PhysRevB.104.064425}%
  \BibitemOpen
  \bibfield  {author} {\bibinfo {author} {\bibfnamefont {L.~O.}\ \bibnamefont
  {Sandberg}}, \bibinfo {author} {\bibfnamefont {R.}~\bibnamefont {Edberg}},
  \bibinfo {author} {\bibfnamefont {I.-M.~B.}\ \bibnamefont {Bakke}}, \bibinfo
  {author} {\bibfnamefont {K.~S.}\ \bibnamefont {Pedersen}}, \bibinfo {author}
  {\bibfnamefont {M.~C.}\ \bibnamefont {Hatnean}}, \bibinfo {author}
  {\bibfnamefont {G.}~\bibnamefont {Balakrishnan}}, \bibinfo {author}
  {\bibfnamefont {L.}~\bibnamefont {Mangin-Thro}}, \bibinfo {author}
  {\bibfnamefont {A.}~\bibnamefont {Wildes}}, \bibinfo {author} {\bibfnamefont
  {B.}~\bibnamefont {F\aa{}k}}, \bibinfo {author} {\bibfnamefont
  {G.}~\bibnamefont {Ehlers}}, \bibinfo {author} {\bibfnamefont
  {G.}~\bibnamefont {Sala}}, \bibinfo {author} {\bibfnamefont {P.}~\bibnamefont
  {Henelius}}, \bibinfo {author} {\bibfnamefont {K.}~\bibnamefont {Lefmann}},\
  and\ \bibinfo {author} {\bibfnamefont {P.~P.}\ \bibnamefont {Deen}},\
  }\bibfield  {title} {\bibinfo {title} {{Emergent magnetic behavior in the
  frustrated ${\mathrm{Yb}}_{3}{\mathrm{Ga}}_{5}{\mathrm{O}}_{12}$ garnet}},\
  }\href {https://doi.org/10.1103/PhysRevB.104.064425} {\bibfield  {journal}
  {\bibinfo  {journal} {Phys. Rev. B}\ }\textbf {\bibinfo {volume} {104}},\
  \bibinfo {pages} {064425} (\bibinfo {year} {2021}{\natexlab{a}})}\BibitemShut
  {NoStop}%
\bibitem [{\citenamefont {Dalmas~de R\'eotier}\ \emph
  {et~al.}(2003)\citenamefont {Dalmas~de R\'eotier}, \citenamefont {Yaouanc},
  \citenamefont {Gubbens}, \citenamefont {Kaiser}, \citenamefont {Baines},\
  and\ \citenamefont {King}}]{PhysRevLett.91.167201}%
  \BibitemOpen
  \bibfield  {author} {\bibinfo {author} {\bibfnamefont {P.}~\bibnamefont
  {Dalmas~de R\'eotier}}, \bibinfo {author} {\bibfnamefont {A.}~\bibnamefont
  {Yaouanc}}, \bibinfo {author} {\bibfnamefont {P.~C.~M.}\ \bibnamefont
  {Gubbens}}, \bibinfo {author} {\bibfnamefont {C.~T.}\ \bibnamefont {Kaiser}},
  \bibinfo {author} {\bibfnamefont {C.}~\bibnamefont {Baines}},\ and\ \bibinfo
  {author} {\bibfnamefont {P.~J.~C.}\ \bibnamefont {King}},\ }\bibfield
  {title} {\bibinfo {title} {{Absence of Magnetic Order in
  ${\mathrm{Y}\mathrm{b}}_{3}{\mathrm{G}\mathrm{a}}_{5}{\mathrm{O}}_{12}$:
  Relation between Phase Transition and Entropy in Geometrically Frustrated
  Materials}},\ }\href {https://doi.org/10.1103/PhysRevLett.91.167201}
  {\bibfield  {journal} {\bibinfo  {journal} {Phys. Rev. Lett.}\ }\textbf
  {\bibinfo {volume} {91}},\ \bibinfo {pages} {167201} (\bibinfo {year}
  {2003})}\BibitemShut {NoStop}%
\bibitem [{\citenamefont {Hodges}\ \emph {et~al.}(2003)\citenamefont {Hodges},
  \citenamefont {Bonville}, \citenamefont {Rams},\ and\ \citenamefont
  {las}}]{Hodges}%
  \BibitemOpen
  \bibfield  {author} {\bibinfo {author} {\bibfnamefont {J.~A.}\ \bibnamefont
  {Hodges}}, \bibinfo {author} {\bibfnamefont {P.}~\bibnamefont {Bonville}},
  \bibinfo {author} {\bibfnamefont {M.}~\bibnamefont {Rams}},\ and\ \bibinfo
  {author} {\bibfnamefont {K.~K.}\ \bibnamefont {las}},\ }\href
  {https://doi.org/10.1088/0953-8984/15/26/313} {\bibfield  {journal} {\bibinfo
   {journal} {Journal of Physics: Condensed Matter}\ }\textbf {\bibinfo
  {volume} {15}},\ \bibinfo {pages} {4631} (\bibinfo {year}
  {2003})}\BibitemShut {NoStop}%
\bibitem [{\citenamefont {Cai}\ \emph {et~al.}(2019)\citenamefont {Cai},
  \citenamefont {Wilson}, \citenamefont {Beare}, \citenamefont {Lygouras},
  \citenamefont {Thomas}, \citenamefont {Yahne}, \citenamefont {Ross},
  \citenamefont {Taddei}, \citenamefont {Sala}, \citenamefont {Dabkowska},
  \citenamefont {Aczel},\ and\ \citenamefont {Luke}}]{PhysRevB.100.184415}%
  \BibitemOpen
  \bibfield  {author} {\bibinfo {author} {\bibfnamefont {Y.}~\bibnamefont
  {Cai}}, \bibinfo {author} {\bibfnamefont {M.~N.}\ \bibnamefont {Wilson}},
  \bibinfo {author} {\bibfnamefont {J.}~\bibnamefont {Beare}}, \bibinfo
  {author} {\bibfnamefont {C.}~\bibnamefont {Lygouras}}, \bibinfo {author}
  {\bibfnamefont {G.}~\bibnamefont {Thomas}}, \bibinfo {author} {\bibfnamefont
  {D.~R.}\ \bibnamefont {Yahne}}, \bibinfo {author} {\bibfnamefont
  {K.}~\bibnamefont {Ross}}, \bibinfo {author} {\bibfnamefont {K.~M.}\
  \bibnamefont {Taddei}}, \bibinfo {author} {\bibfnamefont {G.}~\bibnamefont
  {Sala}}, \bibinfo {author} {\bibfnamefont {H.~A.}\ \bibnamefont {Dabkowska}},
  \bibinfo {author} {\bibfnamefont {A.~A.}\ \bibnamefont {Aczel}},\ and\
  \bibinfo {author} {\bibfnamefont {G.~M.}\ \bibnamefont {Luke}},\ }\bibfield
  {title} {\bibinfo {title} {{Crystal fields and magnetic structure of the
  Ising antiferromagnet
  ${\mathrm{Er}}_{3}{\mathrm{Ga}}_{5}{\mathrm{O}}_{12}$}},\ }\href
  {https://doi.org/10.1103/PhysRevB.100.184415} {\bibfield  {journal} {\bibinfo
   {journal} {Phys. Rev. B}\ }\textbf {\bibinfo {volume} {100}},\ \bibinfo
  {pages} {184415} (\bibinfo {year} {2019})}\BibitemShut {NoStop}%
\bibitem [{\citenamefont {Zhao}\ \emph {et~al.}(2022)\citenamefont {Zhao},
  \citenamefont {Ge}, \citenamefont {Zhou}, \citenamefont {Song}, \citenamefont
  {Yang}, \citenamefont {Li}, \citenamefont {Wang}, \citenamefont {Fu},
  \citenamefont {Zhang}, \citenamefont {Xu}, \citenamefont {Wang},
  \citenamefont {Mei}, \citenamefont {Tong}, \citenamefont {Wu},\ and\
  \citenamefont {Sheng}}]{PhysRevB.105.014441}%
  \BibitemOpen
  \bibfield  {author} {\bibinfo {author} {\bibfnamefont {N.}~\bibnamefont
  {Zhao}}, \bibinfo {author} {\bibfnamefont {H.}~\bibnamefont {Ge}}, \bibinfo
  {author} {\bibfnamefont {L.}~\bibnamefont {Zhou}}, \bibinfo {author}
  {\bibfnamefont {Z.~M.}\ \bibnamefont {Song}}, \bibinfo {author}
  {\bibfnamefont {J.}~\bibnamefont {Yang}}, \bibinfo {author} {\bibfnamefont
  {T.~T.}\ \bibnamefont {Li}}, \bibinfo {author} {\bibfnamefont
  {L.}~\bibnamefont {Wang}}, \bibinfo {author} {\bibfnamefont {Y.}~\bibnamefont
  {Fu}}, \bibinfo {author} {\bibfnamefont {Y.~F.}\ \bibnamefont {Zhang}},
  \bibinfo {author} {\bibfnamefont {J.~B.}\ \bibnamefont {Xu}}, \bibinfo
  {author} {\bibfnamefont {S.~M.}\ \bibnamefont {Wang}}, \bibinfo {author}
  {\bibfnamefont {J.~W.}\ \bibnamefont {Mei}}, \bibinfo {author} {\bibfnamefont
  {X.}~\bibnamefont {Tong}}, \bibinfo {author} {\bibfnamefont {L.~S.}\
  \bibnamefont {Wu}},\ and\ \bibinfo {author} {\bibfnamefont {J.~M.}\
  \bibnamefont {Sheng}},\ }\bibfield  {title} {\bibinfo {title}
  {{Antiferromagnetism and Ising ground states in the rare-earth garnet
  ${\mathrm{Nd}}_{3}{\mathrm{Ga}}_{5}{\mathrm{O}}_{12}$}},\ }\href
  {https://doi.org/10.1103/PhysRevB.105.014441} {\bibfield  {journal} {\bibinfo
   {journal} {Phys. Rev. B}\ }\textbf {\bibinfo {volume} {105}},\ \bibinfo
  {pages} {014441} (\bibinfo {year} {2022})}\BibitemShut {NoStop}%
\bibitem [{\citenamefont {Chilton}\ \emph {et~al.}(2013)\citenamefont
  {Chilton}, \citenamefont {Anderson}, \citenamefont {Turner}, \citenamefont
  {Soncini},\ and\ \citenamefont {Murray}}]{chilton2013phi}%
  \BibitemOpen
  \bibfield  {author} {\bibinfo {author} {\bibfnamefont {N.~F.}\ \bibnamefont
  {Chilton}}, \bibinfo {author} {\bibfnamefont {R.~P.}\ \bibnamefont
  {Anderson}}, \bibinfo {author} {\bibfnamefont {L.~D.}\ \bibnamefont
  {Turner}}, \bibinfo {author} {\bibfnamefont {A.}~\bibnamefont {Soncini}},\
  and\ \bibinfo {author} {\bibfnamefont {K.~S.}\ \bibnamefont {Murray}},\
  }\bibfield  {title} {\bibinfo {title} {{PHI: A powerful new program for the
  analysis of anisotropic monomeric and exchange-coupled polynuclear d-and
  f-block complexes}},\ }\href@noop {} {\bibfield  {journal} {\bibinfo
  {journal} {J. Comput. Chem.}\ }\textbf {\bibinfo {volume} {34}},\ \bibinfo
  {pages} {1164} (\bibinfo {year} {2013})}\BibitemShut {NoStop}%
\bibitem [{\citenamefont
  {Rodríguez-Carvajal}(1993)}]{RODRIGUEZCARVAJAL199355}%
  \BibitemOpen
  \bibfield  {author} {\bibinfo {author} {\bibfnamefont {J.}~\bibnamefont
  {Rodríguez-Carvajal}},\ }\bibfield  {title} {\bibinfo {title} {Recent
  advances in magnetic structure determination by neutron powder diffraction},\
  }\href {https://doi.org/https://doi.org/10.1016/0921-4526(93)90108-I}
  {\bibfield  {journal} {\bibinfo  {journal} {Physica B: Condensed Matter}\
  }\textbf {\bibinfo {volume} {192}},\ \bibinfo {pages} {55} (\bibinfo {year}
  {1993})}\BibitemShut {NoStop}%
\bibitem [{\citenamefont {Momma}\ and\ \citenamefont
  {Izumi}(2011)}]{Momma:db5098}%
  \BibitemOpen
  \bibfield  {author} {\bibinfo {author} {\bibfnamefont {K.}~\bibnamefont
  {Momma}}\ and\ \bibinfo {author} {\bibfnamefont {F.}~\bibnamefont {Izumi}},\
  }\bibfield  {title} {\bibinfo {title} {{{\it VESTA3} for three-dimensional
  visualization of crystal, volumetric and morphology data}},\ }\href
  {https://doi.org/10.1107/S0021889811038970} {\bibfield  {journal} {\bibinfo
  {journal} {Journal of Applied Crystallography}\ }\textbf {\bibinfo {volume}
  {44}},\ \bibinfo {pages} {1272} (\bibinfo {year} {2011})}\BibitemShut
  {NoStop}%
\bibitem [{\citenamefont {Buchanan}\ \emph {et~al.}(1967)\citenamefont
  {Buchanan}, \citenamefont {Wickersheim}, \citenamefont {Pearson},\ and\
  \citenamefont {Herrmann}}]{PhysRev.159.245}%
  \BibitemOpen
  \bibfield  {author} {\bibinfo {author} {\bibfnamefont {R.~A.}\ \bibnamefont
  {Buchanan}}, \bibinfo {author} {\bibfnamefont {K.~A.}\ \bibnamefont
  {Wickersheim}}, \bibinfo {author} {\bibfnamefont {J.~J.}\ \bibnamefont
  {Pearson}},\ and\ \bibinfo {author} {\bibfnamefont {G.~F.}\ \bibnamefont
  {Herrmann}},\ }\bibfield  {title} {\bibinfo {title} {{Energy Levels of
  ${\mathrm{Yb}}^{3+}$ in Gallium and Aluminum Garnets. I. Spectra}},\ }\href
  {https://doi.org/10.1103/PhysRev.159.245} {\bibfield  {journal} {\bibinfo
  {journal} {Phys. Rev.}\ }\textbf {\bibinfo {volume} {159}},\ \bibinfo {pages}
  {245} (\bibinfo {year} {1967})}\BibitemShut {NoStop}%
\bibitem [{\citenamefont {Simonet}\ \emph {et~al.}(2008)\citenamefont
  {Simonet}, \citenamefont {Ballou}, \citenamefont {Robert}, \citenamefont
  {Canals}, \citenamefont {Hippert}, \citenamefont {Bordet}, \citenamefont
  {Lejay}, \citenamefont {Fouquet}, \citenamefont {Ollivier},\ and\
  \citenamefont {Braithwaite}}]{PhysRevLett.100.237204}%
  \BibitemOpen
  \bibfield  {author} {\bibinfo {author} {\bibfnamefont {V.}~\bibnamefont
  {Simonet}}, \bibinfo {author} {\bibfnamefont {R.}~\bibnamefont {Ballou}},
  \bibinfo {author} {\bibfnamefont {J.}~\bibnamefont {Robert}}, \bibinfo
  {author} {\bibfnamefont {B.}~\bibnamefont {Canals}}, \bibinfo {author}
  {\bibfnamefont {F.}~\bibnamefont {Hippert}}, \bibinfo {author} {\bibfnamefont
  {P.}~\bibnamefont {Bordet}}, \bibinfo {author} {\bibfnamefont
  {P.}~\bibnamefont {Lejay}}, \bibinfo {author} {\bibfnamefont
  {P.}~\bibnamefont {Fouquet}}, \bibinfo {author} {\bibfnamefont
  {J.}~\bibnamefont {Ollivier}},\ and\ \bibinfo {author} {\bibfnamefont
  {D.}~\bibnamefont {Braithwaite}},\ }\bibfield  {title} {\bibinfo {title}
  {{Hidden Magnetic Frustration by Quantum Relaxation in Anisotropic Nd
  Langasite}},\ }\href {https://doi.org/10.1103/PhysRevLett.100.237204}
  {\bibfield  {journal} {\bibinfo  {journal} {Phys. Rev. Lett.}\ }\textbf
  {\bibinfo {volume} {100}},\ \bibinfo {pages} {237204} (\bibinfo {year}
  {2008})}\BibitemShut {NoStop}%
\bibitem [{\citenamefont {Shen}\ \emph {et~al.}(2016)\citenamefont {Shen},
  \citenamefont {Li}, \citenamefont {Wo}, \citenamefont {Li}, \citenamefont
  {Shen}, \citenamefont {Pan}, \citenamefont {Wang}, \citenamefont {Walker},
  \citenamefont {Steffens}, \citenamefont {Boehm}, \citenamefont {Hao},
  \citenamefont {Quintero-Castro}, \citenamefont {Harriger}, \citenamefont
  {Frontzek}, \citenamefont {Hao}, \citenamefont {Meng}, \citenamefont {Zhang},
  \citenamefont {Chen},\ and\ \citenamefont {Zhao}}]{Shen2016}%
  \BibitemOpen
  \bibfield  {author} {\bibinfo {author} {\bibfnamefont {Y.}~\bibnamefont
  {Shen}}, \bibinfo {author} {\bibfnamefont {Y.-D.}\ \bibnamefont {Li}},
  \bibinfo {author} {\bibfnamefont {H.}~\bibnamefont {Wo}}, \bibinfo {author}
  {\bibfnamefont {Y.}~\bibnamefont {Li}}, \bibinfo {author} {\bibfnamefont
  {S.}~\bibnamefont {Shen}}, \bibinfo {author} {\bibfnamefont {B.}~\bibnamefont
  {Pan}}, \bibinfo {author} {\bibfnamefont {Q.}~\bibnamefont {Wang}}, \bibinfo
  {author} {\bibfnamefont {H.~C.}\ \bibnamefont {Walker}}, \bibinfo {author}
  {\bibfnamefont {P.}~\bibnamefont {Steffens}}, \bibinfo {author}
  {\bibfnamefont {M.}~\bibnamefont {Boehm}}, \bibinfo {author} {\bibfnamefont
  {Y.}~\bibnamefont {Hao}}, \bibinfo {author} {\bibfnamefont {D.~L.}\
  \bibnamefont {Quintero-Castro}}, \bibinfo {author} {\bibfnamefont {L.~W.}\
  \bibnamefont {Harriger}}, \bibinfo {author} {\bibfnamefont {M.~D.}\
  \bibnamefont {Frontzek}}, \bibinfo {author} {\bibfnamefont {L.}~\bibnamefont
  {Hao}}, \bibinfo {author} {\bibfnamefont {S.}~\bibnamefont {Meng}}, \bibinfo
  {author} {\bibfnamefont {Q.}~\bibnamefont {Zhang}}, \bibinfo {author}
  {\bibfnamefont {G.}~\bibnamefont {Chen}},\ and\ \bibinfo {author}
  {\bibfnamefont {J.}~\bibnamefont {Zhao}},\ }\bibfield  {title} {\bibinfo
  {title} {Evidence for a spinon fermi surface in a triangular-lattice
  quantum-spin-liquid candidate},\ }\href {https://doi.org/10.1038/nature20614}
  {\bibfield  {journal} {\bibinfo  {journal} {Nature}\ }\textbf {\bibinfo
  {volume} {540}},\ \bibinfo {pages} {559} (\bibinfo {year}
  {2016})}\BibitemShut {NoStop}%
\bibitem [{\citenamefont {Lhotel}\ \emph {et~al.}(2021)\citenamefont {Lhotel},
  \citenamefont {Mangin-Thro}, \citenamefont {Ressouche}, \citenamefont
  {Steffens}, \citenamefont {Bichaud}, \citenamefont {Knebel}, \citenamefont
  {Brison}, \citenamefont {Marin}, \citenamefont {Raymond},\ and\ \citenamefont
  {Zhitomirsky}}]{PhysRevB.104.024427}%
  \BibitemOpen
  \bibfield  {author} {\bibinfo {author} {\bibfnamefont {E.}~\bibnamefont
  {Lhotel}}, \bibinfo {author} {\bibfnamefont {L.}~\bibnamefont {Mangin-Thro}},
  \bibinfo {author} {\bibfnamefont {E.}~\bibnamefont {Ressouche}}, \bibinfo
  {author} {\bibfnamefont {P.}~\bibnamefont {Steffens}}, \bibinfo {author}
  {\bibfnamefont {E.}~\bibnamefont {Bichaud}}, \bibinfo {author} {\bibfnamefont
  {G.}~\bibnamefont {Knebel}}, \bibinfo {author} {\bibfnamefont {J.-P.}\
  \bibnamefont {Brison}}, \bibinfo {author} {\bibfnamefont {C.}~\bibnamefont
  {Marin}}, \bibinfo {author} {\bibfnamefont {S.}~\bibnamefont {Raymond}},\
  and\ \bibinfo {author} {\bibfnamefont {M.~E.}\ \bibnamefont {Zhitomirsky}},\
  }\bibfield  {title} {\bibinfo {title} {{Spin dynamics of the quantum dipolar
  magnet ${\mathrm{Yb}}_{3}{\mathrm{Ga}}_{5}{\mathrm{O}}_{12}$ in an external
  field}},\ }\href {https://doi.org/10.1103/PhysRevB.104.024427} {\bibfield
  {journal} {\bibinfo  {journal} {Phys. Rev. B}\ }\textbf {\bibinfo {volume}
  {104}},\ \bibinfo {pages} {024427} (\bibinfo {year} {2021})}\BibitemShut
  {NoStop}%
\bibitem [{\citenamefont {Jensen}\ and\ \citenamefont
  {Mackintosh}(1991)}]{jense1991rare}%
  \BibitemOpen
  \bibfield  {author} {\bibinfo {author} {\bibfnamefont {J.}~\bibnamefont
  {Jensen}}\ and\ \bibinfo {author} {\bibfnamefont {A.~R.}\ \bibnamefont
  {Mackintosh}},\ }\href@noop {} {\emph {\bibinfo {title} {Rare Earth
  Magnetism: Structures and Excitations}}}\ (\bibinfo  {publisher} {Oxford
  Science Publications. Clarendon Press},\ \bibinfo {year} {1991})\BibitemShut
  {NoStop}%
\bibitem [{\citenamefont {Sosin}\ \emph {et~al.}(2022)\citenamefont {Sosin},
  \citenamefont {Iafarova}, \citenamefont {Romanova}, \citenamefont {Morozov},
  \citenamefont {Korableva}, \citenamefont {Batulin}, \citenamefont
  {Zhitomirsky},\ and\ \citenamefont {Glazkov}}]{Sosin2022}%
  \BibitemOpen
  \bibfield  {author} {\bibinfo {author} {\bibfnamefont {S.~S.}\ \bibnamefont
  {Sosin}}, \bibinfo {author} {\bibfnamefont {A.~F.}\ \bibnamefont {Iafarova}},
  \bibinfo {author} {\bibfnamefont {I.~V.}\ \bibnamefont {Romanova}}, \bibinfo
  {author} {\bibfnamefont {O.~A.}\ \bibnamefont {Morozov}}, \bibinfo {author}
  {\bibfnamefont {S.~L.}\ \bibnamefont {Korableva}}, \bibinfo {author}
  {\bibfnamefont {R.~G.}\ \bibnamefont {Batulin}}, \bibinfo {author}
  {\bibfnamefont {M.}~\bibnamefont {Zhitomirsky}},\ and\ \bibinfo {author}
  {\bibfnamefont {V.~N.}\ \bibnamefont {Glazkov}},\ }\bibfield  {title}
  {\bibinfo {title} {{Microscopic Spin Hamiltonian for a Dipolar Heisenberg
  Magnet LiGdF4 from EPR Measurements}},\ }\href
  {https://doi.org/10.1134/S0021364022602299} {\bibfield  {journal} {\bibinfo
  {journal} {JETP Letters}\ }\textbf {\bibinfo {volume} {116}},\ \bibinfo
  {pages} {771} (\bibinfo {year} {2022})}\BibitemShut {NoStop}%
\bibitem [{\citenamefont {Hamilton}\ \emph {et~al.}(2014)\citenamefont
  {Hamilton}, \citenamefont {aand S~E~Rowley},\ and\ \citenamefont
  {Dutton}}]{Sackville_Hamilton_2014}%
  \BibitemOpen
  \bibfield  {author} {\bibinfo {author} {\bibfnamefont {A.~C.~S.}\
  \bibnamefont {Hamilton}}, \bibinfo {author} {\bibfnamefont {G.~I.~L.}\
  \bibnamefont {aand S~E~Rowley}},\ and\ \bibinfo {author} {\bibfnamefont
  {S.~E.}\ \bibnamefont {Dutton}},\ }\bibfield  {title} {\bibinfo {title}
  {{Enhancement of the magnetocaloric effect driven by changes in the crystal
  structure of Al-doped GGG, Gd$_3$Ga$_{5−x}$Al$_x$O$_{12}$ (0
  $\leq$x$\leq$5)}},\ }\href {https://doi.org/10.1088/0953-8984/26/11/116001}
  {\bibfield  {journal} {\bibinfo  {journal} {Journal of Physics: Condensed
  Matter}\ }\textbf {\bibinfo {volume} {26}},\ \bibinfo {pages} {116001}
  (\bibinfo {year} {2014})}\BibitemShut {NoStop}%
\bibitem [{\citenamefont {Luo}\ \emph {et~al.}(2018)\citenamefont {Luo},
  \citenamefont {Lake}, \citenamefont {Mei},\ and\ \citenamefont
  {Starykh}}]{PhysRevLett.120.037204}%
  \BibitemOpen
  \bibfield  {author} {\bibinfo {author} {\bibfnamefont {Z.-X.}\ \bibnamefont
  {Luo}}, \bibinfo {author} {\bibfnamefont {E.}~\bibnamefont {Lake}}, \bibinfo
  {author} {\bibfnamefont {J.-W.}\ \bibnamefont {Mei}},\ and\ \bibinfo {author}
  {\bibfnamefont {O.~A.}\ \bibnamefont {Starykh}},\ }\bibfield  {title}
  {\bibinfo {title} {Spinon magnetic resonance of quantum spin liquids},\
  }\href {https://doi.org/10.1103/PhysRevLett.120.037204} {\bibfield  {journal}
  {\bibinfo  {journal} {Phys. Rev. Lett.}\ }\textbf {\bibinfo {volume} {120}},\
  \bibinfo {pages} {037204} (\bibinfo {year} {2018})}\BibitemShut {NoStop}%
\bibitem [{\citenamefont {Somesh}\ \emph {et~al.}(2023)\citenamefont {Somesh},
  \citenamefont {Islam}, \citenamefont {Mohanty}, \citenamefont {Simutis},
  \citenamefont {Guguchia}, \citenamefont {Wang}, \citenamefont
  {Sichelschmidt}, \citenamefont {Baenitz},\ and\ \citenamefont
  {Nath}}]{PhysRevB.107.064421}%
  \BibitemOpen
  \bibfield  {author} {\bibinfo {author} {\bibfnamefont {K.}~\bibnamefont
  {Somesh}}, \bibinfo {author} {\bibfnamefont {S.~S.}\ \bibnamefont {Islam}},
  \bibinfo {author} {\bibfnamefont {S.}~\bibnamefont {Mohanty}}, \bibinfo
  {author} {\bibfnamefont {G.}~\bibnamefont {Simutis}}, \bibinfo {author}
  {\bibfnamefont {Z.}~\bibnamefont {Guguchia}}, \bibinfo {author}
  {\bibfnamefont {C.}~\bibnamefont {Wang}}, \bibinfo {author} {\bibfnamefont
  {J.}~\bibnamefont {Sichelschmidt}}, \bibinfo {author} {\bibfnamefont
  {M.}~\bibnamefont {Baenitz}},\ and\ \bibinfo {author} {\bibfnamefont
  {R.}~\bibnamefont {Nath}},\ }\bibfield  {title} {\bibinfo {title} {{Absence
  of magnetic order and emergence of unconventional fluctuations in the
  ${J}_{\mathrm{eff}}=\frac{1}{2}$ triangular-lattice antiferromagnet
  ${\mathrm{YbBO}}_{3}$}},\ }\href
  {https://doi.org/10.1103/PhysRevB.107.064421} {\bibfield  {journal} {\bibinfo
   {journal} {Phys. Rev. B}\ }\textbf {\bibinfo {volume} {107}},\ \bibinfo
  {pages} {064421} (\bibinfo {year} {2023})}\BibitemShut {NoStop}%
\bibitem [{\citenamefont {Benner}\ \emph {et~al.}(1983)\citenamefont {Benner},
  \citenamefont {Brodehl}, \citenamefont {Seitz},\ and\ \citenamefont
  {Wiese}}]{H_Benner_1983}%
  \BibitemOpen
  \bibfield  {author} {\bibinfo {author} {\bibfnamefont {H.}~\bibnamefont
  {Benner}}, \bibinfo {author} {\bibfnamefont {M.}~\bibnamefont {Brodehl}},
  \bibinfo {author} {\bibfnamefont {H.}~\bibnamefont {Seitz}},\ and\ \bibinfo
  {author} {\bibfnamefont {J.}~\bibnamefont {Wiese}},\ }\bibfield  {title}
  {\bibinfo {title} {Influence of nondiagonal dynamic susceptibility on the epr
  signal of heisenberg magnets},\ }\href
  {https://doi.org/10.1088/0022-3719/16/31/015} {\bibfield  {journal} {\bibinfo
   {journal} {Journal of Physics C: Solid State Physics}\ }\textbf {\bibinfo
  {volume} {16}},\ \bibinfo {pages} {6011} (\bibinfo {year}
  {1983})}\BibitemShut {NoStop}%
\bibitem [{\citenamefont {Tari}(2003)}]{tari2003specific}%
  \BibitemOpen
  \bibfield  {author} {\bibinfo {author} {\bibfnamefont {A.}~\bibnamefont
  {Tari}},\ }\href@noop {} {\emph {\bibinfo {title} {The specific heat of
  matter at low temperatures}}}\ (\bibinfo  {publisher} {World Scientific},\
  \bibinfo {year} {2003})\BibitemShut {NoStop}%
\bibitem [{\citenamefont {Gopal}(2012)}]{gopal2012specific}%
  \BibitemOpen
  \bibfield  {author} {\bibinfo {author} {\bibfnamefont {E.}~\bibnamefont
  {Gopal}},\ }\href@noop {} {\emph {\bibinfo {title} {Specific heats at low
  temperatures}}}\ (\bibinfo  {publisher} {Springer Science \& Business
  Media},\ \bibinfo {year} {2012})\BibitemShut {NoStop}%
\bibitem [{\citenamefont {Jiang}\ \emph {et~al.}(2022)\citenamefont {Jiang},
  \citenamefont {Yang}, \citenamefont {Gao}, \citenamefont {Wan}, \citenamefont
  {Zhu}, \citenamefont {Shiroka}, \citenamefont {Chen}, \citenamefont {Wu},
  \citenamefont {Li}, \citenamefont {Jiao}, \citenamefont {Chen}, \citenamefont
  {Bao}, \citenamefont {Tian},\ and\ \citenamefont
  {Shu}}]{PhysRevB.106.014409}%
  \BibitemOpen
  \bibfield  {author} {\bibinfo {author} {\bibfnamefont {C.~Y.}\ \bibnamefont
  {Jiang}}, \bibinfo {author} {\bibfnamefont {Y.~X.}\ \bibnamefont {Yang}},
  \bibinfo {author} {\bibfnamefont {Y.~X.}\ \bibnamefont {Gao}}, \bibinfo
  {author} {\bibfnamefont {Z.~T.}\ \bibnamefont {Wan}}, \bibinfo {author}
  {\bibfnamefont {Z.~H.}\ \bibnamefont {Zhu}}, \bibinfo {author} {\bibfnamefont
  {T.}~\bibnamefont {Shiroka}}, \bibinfo {author} {\bibfnamefont {C.~S.}\
  \bibnamefont {Chen}}, \bibinfo {author} {\bibfnamefont {Q.}~\bibnamefont
  {Wu}}, \bibinfo {author} {\bibfnamefont {X.}~\bibnamefont {Li}}, \bibinfo
  {author} {\bibfnamefont {J.~C.}\ \bibnamefont {Jiao}}, \bibinfo {author}
  {\bibfnamefont {K.~W.}\ \bibnamefont {Chen}}, \bibinfo {author}
  {\bibfnamefont {Y.}~\bibnamefont {Bao}}, \bibinfo {author} {\bibfnamefont
  {Z.~M.}\ \bibnamefont {Tian}},\ and\ \bibinfo {author} {\bibfnamefont
  {L.}~\bibnamefont {Shu}},\ }\bibfield  {title} {\bibinfo {title} {Spin
  excitations in the quantum dipolar magnet
  $\mathrm{Yb}{({\mathrm{BaBO}}_{3})}_{3}$},\ }\href
  {https://doi.org/10.1103/PhysRevB.106.014409} {\bibfield  {journal} {\bibinfo
   {journal} {Phys. Rev. B}\ }\textbf {\bibinfo {volume} {106}},\ \bibinfo
  {pages} {014409} (\bibinfo {year} {2022})}\BibitemShut {NoStop}%
\bibitem [{\citenamefont {Filippi}\ \emph {et~al.}(1980)\citenamefont
  {Filippi}, \citenamefont {Lasjaunias}, \citenamefont {Hebral}, \citenamefont
  {Rossat-Mignod},\ and\ \citenamefont {Tcheou}}]{JFilippi1980}%
  \BibitemOpen
  \bibfield  {author} {\bibinfo {author} {\bibfnamefont {J.}~\bibnamefont
  {Filippi}}, \bibinfo {author} {\bibfnamefont {J.~C.}\ \bibnamefont
  {Lasjaunias}}, \bibinfo {author} {\bibfnamefont {B.}~\bibnamefont {Hebral}},
  \bibinfo {author} {\bibfnamefont {J.}~\bibnamefont {Rossat-Mignod}},\ and\
  \bibinfo {author} {\bibfnamefont {F.}~\bibnamefont {Tcheou}},\ }\bibfield
  {title} {\bibinfo {title} {{Magnetic properties of ytterbium gallium garnet
  between 44 mK and 4 K}},\ }\href {https://doi.org/10.1088/0022-3719/13/7/012}
  {\bibfield  {journal} {\bibinfo  {journal} {Journal of Physics C: Solid State
  Physics}\ }\textbf {\bibinfo {volume} {13}},\ \bibinfo {pages} {1277}
  (\bibinfo {year} {1980})}\BibitemShut {NoStop}%
\bibitem [{\citenamefont {Lounasmaa}(1962)}]{PhysRev.128.1136}%
  \BibitemOpen
  \bibfield  {author} {\bibinfo {author} {\bibfnamefont {O.~V.}\ \bibnamefont
  {Lounasmaa}},\ }\bibfield  {title} {\bibinfo {title} {{Specific Heat of
  Holmium Metal between 0.38 and 4.2\ifmmode^\circ\else\textdegree\fi{}K}},\
  }\href {https://doi.org/10.1103/PhysRev.128.1136} {\bibfield  {journal}
  {\bibinfo  {journal} {Phys. Rev.}\ }\textbf {\bibinfo {volume} {128}},\
  \bibinfo {pages} {1136} (\bibinfo {year} {1962})}\BibitemShut {NoStop}%
\bibitem [{\citenamefont {Takeda}\ \emph {et~al.}(2008)\citenamefont {Takeda},
  \citenamefont {Duc~Dung}, \citenamefont {Nakano}, \citenamefont {Ishikura},
  \citenamefont {Ikeda}, \citenamefont {D.~Matsuda}, \citenamefont {Yamamoto},
  \citenamefont {Haga}, \citenamefont {Takeuchi}, \citenamefont {Settai},\ and\
  \citenamefont {Ōnuki}}]{doi:10.1143/JPSJ.77.104710}%
  \BibitemOpen
  \bibfield  {author} {\bibinfo {author} {\bibfnamefont {Y.}~\bibnamefont
  {Takeda}}, \bibinfo {author} {\bibfnamefont {N.}~\bibnamefont {Duc~Dung}},
  \bibinfo {author} {\bibfnamefont {Y.}~\bibnamefont {Nakano}}, \bibinfo
  {author} {\bibfnamefont {T.}~\bibnamefont {Ishikura}}, \bibinfo {author}
  {\bibfnamefont {S.}~\bibnamefont {Ikeda}}, \bibinfo {author} {\bibfnamefont
  {T.}~\bibnamefont {D.~Matsuda}}, \bibinfo {author} {\bibfnamefont
  {E.}~\bibnamefont {Yamamoto}}, \bibinfo {author} {\bibfnamefont
  {Y.}~\bibnamefont {Haga}}, \bibinfo {author} {\bibfnamefont {T.}~\bibnamefont
  {Takeuchi}}, \bibinfo {author} {\bibfnamefont {R.}~\bibnamefont {Settai}},\
  and\ \bibinfo {author} {\bibfnamefont {Y.}~\bibnamefont {Ōnuki}},\
  }\bibfield  {title} {\bibinfo {title} {{Calorimetric Study in Single
  Crystalline RCu$_2$Si$_2$ (R: Rare Earth)}},\ }\href
  {https://doi.org/10.1143/JPSJ.77.104710} {\bibfield  {journal} {\bibinfo
  {journal} {Journal of the Physical Society of Japan}\ }\textbf {\bibinfo
  {volume} {77}},\ \bibinfo {pages} {104710} (\bibinfo {year}
  {2008})}\BibitemShut {NoStop}%
\bibitem [{\citenamefont {Kundu}\ \emph {et~al.}(2020)\citenamefont {Kundu},
  \citenamefont {Shahee}, \citenamefont {Chakraborty}, \citenamefont {Ranjith},
  \citenamefont {Koo}, \citenamefont {Sichelschmidt}, \citenamefont {Telling},
  \citenamefont {Biswas}, \citenamefont {Baenitz}, \citenamefont {Dasgupta},
  \citenamefont {Pujari},\ and\ \citenamefont
  {Mahajan}}]{PhysRevLett.125.267202}%
  \BibitemOpen
  \bibfield  {author} {\bibinfo {author} {\bibfnamefont {S.}~\bibnamefont
  {Kundu}}, \bibinfo {author} {\bibfnamefont {A.}~\bibnamefont {Shahee}},
  \bibinfo {author} {\bibfnamefont {A.}~\bibnamefont {Chakraborty}}, \bibinfo
  {author} {\bibfnamefont {K.~M.}\ \bibnamefont {Ranjith}}, \bibinfo {author}
  {\bibfnamefont {B.}~\bibnamefont {Koo}}, \bibinfo {author} {\bibfnamefont
  {J.}~\bibnamefont {Sichelschmidt}}, \bibinfo {author} {\bibfnamefont
  {M.~T.~F.}\ \bibnamefont {Telling}}, \bibinfo {author} {\bibfnamefont
  {P.~K.}\ \bibnamefont {Biswas}}, \bibinfo {author} {\bibfnamefont
  {M.}~\bibnamefont {Baenitz}}, \bibinfo {author} {\bibfnamefont
  {I.}~\bibnamefont {Dasgupta}}, \bibinfo {author} {\bibfnamefont
  {S.}~\bibnamefont {Pujari}},\ and\ \bibinfo {author} {\bibfnamefont {A.~V.}\
  \bibnamefont {Mahajan}},\ }\bibfield  {title} {\bibinfo {title} {{Gapless
  Quantum Spin Liquid in the Triangular System
  ${\mathrm{Sr}}_{3}{\mathrm{CuSb}}_{2}{\mathrm{O}}_{9}$}},\ }\href
  {https://doi.org/10.1103/PhysRevLett.125.267202} {\bibfield  {journal}
  {\bibinfo  {journal} {Phys. Rev. Lett.}\ }\textbf {\bibinfo {volume} {125}},\
  \bibinfo {pages} {267202} (\bibinfo {year} {2020})}\BibitemShut {NoStop}%
\bibitem [{\citenamefont {Dey}\ \emph {et~al.}(2012)\citenamefont {Dey},
  \citenamefont {Mahajan}, \citenamefont {Khuntia}, \citenamefont {Baenitz},
  \citenamefont {Koteswararao},\ and\ \citenamefont
  {Chou}}]{PhysRevB.86.140405}%
  \BibitemOpen
  \bibfield  {author} {\bibinfo {author} {\bibfnamefont {T.}~\bibnamefont
  {Dey}}, \bibinfo {author} {\bibfnamefont {A.~V.}\ \bibnamefont {Mahajan}},
  \bibinfo {author} {\bibfnamefont {P.}~\bibnamefont {Khuntia}}, \bibinfo
  {author} {\bibfnamefont {M.}~\bibnamefont {Baenitz}}, \bibinfo {author}
  {\bibfnamefont {B.}~\bibnamefont {Koteswararao}},\ and\ \bibinfo {author}
  {\bibfnamefont {F.~C.}\ \bibnamefont {Chou}},\ }\bibfield  {title} {\bibinfo
  {title} {{Spin-liquid behavior in ${J}_{\mathrm{eff}}=\frac{1}{2}$ triangular
  lattice compound {Ba$_3$IrTi$_2$O$_9$}}},\ }\href
  {https://doi.org/10.1103/PhysRevB.86.140405} {\bibfield  {journal} {\bibinfo
  {journal} {Phys. Rev. B}\ }\textbf {\bibinfo {volume} {86}},\ \bibinfo
  {pages} {140405} (\bibinfo {year} {2012})}\BibitemShut {NoStop}%
\bibitem [{\citenamefont {de~Vries}\ \emph {et~al.}(2008)\citenamefont
  {de~Vries}, \citenamefont {Kamenev}, \citenamefont {Kockelmann},
  \citenamefont {Sanchez-Benitez},\ and\ \citenamefont
  {Harrison}}]{PhysRevLett.100.157205}%
  \BibitemOpen
  \bibfield  {author} {\bibinfo {author} {\bibfnamefont {M.~A.}\ \bibnamefont
  {de~Vries}}, \bibinfo {author} {\bibfnamefont {K.~V.}\ \bibnamefont
  {Kamenev}}, \bibinfo {author} {\bibfnamefont {W.~A.}\ \bibnamefont
  {Kockelmann}}, \bibinfo {author} {\bibfnamefont {J.}~\bibnamefont
  {Sanchez-Benitez}},\ and\ \bibinfo {author} {\bibfnamefont {A.}~\bibnamefont
  {Harrison}},\ }\bibfield  {title} {\bibinfo {title} {{Magnetic Ground State
  of an Experimental $S=1/2$ Kagome Antiferromagnet}},\ }\href
  {https://doi.org/10.1103/PhysRevLett.100.157205} {\bibfield  {journal}
  {\bibinfo  {journal} {Phys. Rev. Lett.}\ }\textbf {\bibinfo {volume} {100}},\
  \bibinfo {pages} {157205} (\bibinfo {year} {2008})}\BibitemShut {NoStop}%
\bibitem [{\citenamefont {Paddison}\ \emph {et~al.}(2017)\citenamefont
  {Paddison}, \citenamefont {Daum}, \citenamefont {Dun}, \citenamefont
  {Ehlers}, \citenamefont {Liu}, \citenamefont {Stone}, \citenamefont {Zhou},\
  and\ \citenamefont {Mourigal}}]{Paddison2017}%
  \BibitemOpen
  \bibfield  {author} {\bibinfo {author} {\bibfnamefont {J.~A.~M.}\
  \bibnamefont {Paddison}}, \bibinfo {author} {\bibfnamefont {M.}~\bibnamefont
  {Daum}}, \bibinfo {author} {\bibfnamefont {Z.}~\bibnamefont {Dun}}, \bibinfo
  {author} {\bibfnamefont {G.}~\bibnamefont {Ehlers}}, \bibinfo {author}
  {\bibfnamefont {Y.}~\bibnamefont {Liu}}, \bibinfo {author} {\bibfnamefont
  {M.}~\bibnamefont {Stone}}, \bibinfo {author} {\bibfnamefont
  {H.}~\bibnamefont {Zhou}},\ and\ \bibinfo {author} {\bibfnamefont
  {M.}~\bibnamefont {Mourigal}},\ }\bibfield  {title} {\bibinfo {title}
  {{Continuous excitations of the triangular-lattice quantum spin liquid
  YbMgGaO$_4$}},\ }\href {https://doi.org/10.1038/nphys3971} {\bibfield
  {journal} {\bibinfo  {journal} {Nature Physics}\ }\textbf {\bibinfo {volume}
  {13}},\ \bibinfo {pages} {117} (\bibinfo {year} {2017})}\BibitemShut
  {NoStop}%
\bibitem [{\citenamefont {Baenitz}\ \emph {et~al.}(2018)\citenamefont
  {Baenitz}, \citenamefont {Schlender}, \citenamefont {Sichelschmidt},
  \citenamefont {Onykiienko}, \citenamefont {Zangeneh}, \citenamefont
  {Ranjith}, \citenamefont {Sarkar}, \citenamefont {Hozoi}, \citenamefont
  {Walker}, \citenamefont {Orain}, \citenamefont {Yasuoka}, \citenamefont
  {van~den Brink}, \citenamefont {Klauss}, \citenamefont {Inosov},\ and\
  \citenamefont {Doert}}]{PhysRevB.98.220409}%
  \BibitemOpen
  \bibfield  {author} {\bibinfo {author} {\bibfnamefont {M.}~\bibnamefont
  {Baenitz}}, \bibinfo {author} {\bibfnamefont {P.}~\bibnamefont {Schlender}},
  \bibinfo {author} {\bibfnamefont {J.}~\bibnamefont {Sichelschmidt}}, \bibinfo
  {author} {\bibfnamefont {Y.~A.}\ \bibnamefont {Onykiienko}}, \bibinfo
  {author} {\bibfnamefont {Z.}~\bibnamefont {Zangeneh}}, \bibinfo {author}
  {\bibfnamefont {K.~M.}\ \bibnamefont {Ranjith}}, \bibinfo {author}
  {\bibfnamefont {R.}~\bibnamefont {Sarkar}}, \bibinfo {author} {\bibfnamefont
  {L.}~\bibnamefont {Hozoi}}, \bibinfo {author} {\bibfnamefont {H.~C.}\
  \bibnamefont {Walker}}, \bibinfo {author} {\bibfnamefont {J.-C.}\
  \bibnamefont {Orain}}, \bibinfo {author} {\bibfnamefont {H.}~\bibnamefont
  {Yasuoka}}, \bibinfo {author} {\bibfnamefont {J.}~\bibnamefont {van~den
  Brink}}, \bibinfo {author} {\bibfnamefont {H.~H.}\ \bibnamefont {Klauss}},
  \bibinfo {author} {\bibfnamefont {D.~S.}\ \bibnamefont {Inosov}},\ and\
  \bibinfo {author} {\bibfnamefont {T.}~\bibnamefont {Doert}},\ }\bibfield
  {title} {\bibinfo {title} {{${\mathrm{NaYbS}}_{2}$: A planar
  spin-$\frac{1}{2}$ triangular-lattice magnet and putative spin liquid}},\
  }\href {https://doi.org/10.1103/PhysRevB.98.220409} {\bibfield  {journal}
  {\bibinfo  {journal} {Phys. Rev. B}\ }\textbf {\bibinfo {volume} {98}},\
  \bibinfo {pages} {220409} (\bibinfo {year} {2018})}\BibitemShut {NoStop}%
\bibitem [{\citenamefont {Yamashita}\ \emph {et~al.}(2008)\citenamefont
  {Yamashita}, \citenamefont {Nakazawa}, \citenamefont {Oguni}, \citenamefont
  {Oshima}, \citenamefont {Nojiri}, \citenamefont {Shimizu}, \citenamefont
  {Miyagawa},\ and\ \citenamefont {Kanoda}}]{Yamashita2008}%
  \BibitemOpen
  \bibfield  {author} {\bibinfo {author} {\bibfnamefont {S.}~\bibnamefont
  {Yamashita}}, \bibinfo {author} {\bibfnamefont {Y.}~\bibnamefont {Nakazawa}},
  \bibinfo {author} {\bibfnamefont {M.}~\bibnamefont {Oguni}}, \bibinfo
  {author} {\bibfnamefont {Y.}~\bibnamefont {Oshima}}, \bibinfo {author}
  {\bibfnamefont {H.}~\bibnamefont {Nojiri}}, \bibinfo {author} {\bibfnamefont
  {Y.}~\bibnamefont {Shimizu}}, \bibinfo {author} {\bibfnamefont
  {K.}~\bibnamefont {Miyagawa}},\ and\ \bibinfo {author} {\bibfnamefont
  {K.}~\bibnamefont {Kanoda}},\ }\bibfield  {title} {\bibinfo {title}
  {Thermodynamic properties of a spin-1/2 spin-liquid state in a $\kappa$-type
  organic salt},\ }\href {https://doi.org/10.1038/nphys942} {\bibfield
  {journal} {\bibinfo  {journal} {Nature Physics}\ }\textbf {\bibinfo {volume}
  {4}},\ \bibinfo {pages} {459} (\bibinfo {year} {2008})}\BibitemShut {NoStop}%
\bibitem [{\citenamefont {Yamashita}\ \emph {et~al.}(2011)\citenamefont
  {Yamashita}, \citenamefont {Yamamoto}, \citenamefont {Nakazawa},
  \citenamefont {Tamura},\ and\ \citenamefont {Kato}}]{Yamashita2011}%
  \BibitemOpen
  \bibfield  {author} {\bibinfo {author} {\bibfnamefont {S.}~\bibnamefont
  {Yamashita}}, \bibinfo {author} {\bibfnamefont {T.}~\bibnamefont {Yamamoto}},
  \bibinfo {author} {\bibfnamefont {Y.}~\bibnamefont {Nakazawa}}, \bibinfo
  {author} {\bibfnamefont {M.}~\bibnamefont {Tamura}},\ and\ \bibinfo {author}
  {\bibfnamefont {R.}~\bibnamefont {Kato}},\ }\bibfield  {title} {\bibinfo
  {title} {Gapless spin liquid of an organic triangular compound evidenced by
  thermodynamic measurements},\ }\href {https://doi.org/10.1038/ncomms1274}
  {\bibfield  {journal} {\bibinfo  {journal} {Nature Communications}\ }\textbf
  {\bibinfo {volume} {2}},\ \bibinfo {pages} {275} (\bibinfo {year}
  {2011})}\BibitemShut {NoStop}%
\bibitem [{\citenamefont {Stevens}(1952)}]{stevens1952matrix}%
  \BibitemOpen
  \bibfield  {author} {\bibinfo {author} {\bibfnamefont {K.~W.~H.}\
  \bibnamefont {Stevens}},\ }\bibfield  {title} {\bibinfo {title} {{Matrix
  elements and operator equivalents connected with the magnetic properties of
  rare earth ions}},\ }\href {https://doi.org/10.1088/0370-1298/65/3/308}
  {\bibfield  {journal} {\bibinfo  {journal} {Proc. Phys. Soc.}\ }\textbf
  {\bibinfo {volume} {65}},\ \bibinfo {pages} {209} (\bibinfo {year}
  {1952})}\BibitemShut {NoStop}%
\bibitem [{\citenamefont {Sandberg}\ \emph
  {et~al.}(2021{\natexlab{b}})\citenamefont {Sandberg}, \citenamefont {Edberg},
  \citenamefont {Bakke}, \citenamefont {Pedersen}, \citenamefont {Hatnean},
  \citenamefont {Balakrishnan}, \citenamefont {Mangin-Thro}, \citenamefont
  {Wildes}, \citenamefont {F{\aa}k}, \citenamefont {Ehlers} \emph
  {et~al.}}]{sandberg2021emergent}%
  \BibitemOpen
  \bibfield  {author} {\bibinfo {author} {\bibfnamefont {L.~{\O}.}\
  \bibnamefont {Sandberg}}, \bibinfo {author} {\bibfnamefont {R.}~\bibnamefont
  {Edberg}}, \bibinfo {author} {\bibfnamefont {I.-M.~B.}\ \bibnamefont
  {Bakke}}, \bibinfo {author} {\bibfnamefont {K.~S.}\ \bibnamefont {Pedersen}},
  \bibinfo {author} {\bibfnamefont {M.~C.}\ \bibnamefont {Hatnean}}, \bibinfo
  {author} {\bibfnamefont {G.}~\bibnamefont {Balakrishnan}}, \bibinfo {author}
  {\bibfnamefont {L.}~\bibnamefont {Mangin-Thro}}, \bibinfo {author}
  {\bibfnamefont {A.}~\bibnamefont {Wildes}}, \bibinfo {author} {\bibfnamefont
  {B.}~\bibnamefont {F{\aa}k}}, \bibinfo {author} {\bibfnamefont
  {G.}~\bibnamefont {Ehlers}}, \emph {et~al.},\ }\bibfield  {title} {\bibinfo
  {title} {{Emergent magnetic behavior in the frustrated Yb$_3$Ga$_5$O$_{12}$
  garnet}},\ }\href {https://doi.org/10.1103/PhysRevB.104.064425} {\bibfield
  {journal} {\bibinfo  {journal} {Phys. Rev. B}\ }\textbf {\bibinfo {volume}
  {104}},\ \bibinfo {pages} {064425} (\bibinfo {year}
  {2021}{\natexlab{b}})}\BibitemShut {NoStop}%
\end{thebibliography}%
\end{document}